\documentclass[sigconf,authorversion]{acmart}

\usepackage{flushend} % balance last page
\usepackage{balance}

\usepackage{placeins} % -> use \FloatBarrier

\usepackage{stfloats}

\usepackage[nopostdot,acronym,shortcuts,nonumberlist]{glossaries}
\newacronym{(S+N)/NR}{(S+N)/NR}{signal-plus-noise-to-noise ratio}
\newacronym{ADC}{ADC}{analog-to-digital converter}
\newacronym{BCI}{BCI}{bulk current injection}
\newacronym{CAN}{CAN}{control area network}
\newacronym{CRT}{CRT}{cathode ray tube}
\newacronym{CSMACD}{CSMA/CD}{carrier sense multiple access with collision detection}
\newacronym{CoE}{CoE}{CAN over Ethernet}
\newacronym{DAC}{DAC}{digital-to-analog converter}
\newacronym{DoS}{DoS}{Denial of Service}
\newacronym{EMR}{EMR}{electromagnetic radiation}
\newacronym{FCS}{FCS}{frame check sum}
\newacronym{FEC}{FEC}{forward error correction}
\newacronym{FPGA}{FPGA}{field-programmable gate array}
\newacronym{ICMP}{ICMP}{Internet control message protocol}
\newacronym{IEC}{IEC}{International Electrotechnical Commission}
\newacronym{MAC}{MAC}{Medium Access Control}
\newacronym{NIC}{NIC}{network interface card}
\newacronym{NSA}{NSA}{National Security Agency}
\newacronym{PA}{PA}{power amplifier}
\newacronym{PC}{PC}{personal computer}
\newacronym{RF}{RF}{radio frequency}
\newacronym{SDR}{SDR}{software-defined radio}
\newacronym{SFD}{SFD}{start frame delimiter}
\newacronym{SNR}{SNR}{signal-to-noise ratio}
\newacronym{USRP}{USRP}{Universal Software Radio Peripheral}
\newacronym{SSP}{SSP}{Secure Simple Pairing}
\newacronym{USB}{USB}{Universal Serial Bus}
\newacronym{TLV}{TLV}{Type Length Value}
\newacronym{L2CAP}{L2CAP}{Logical Link Control and Adaptation Protocol}
\newacronym{AOP}{AOP}{Always-On Processor}
\newacronym{ACL}{ACL}{Asynchronous Connec\-tion-Less}
\newacronym{BLE}{BLE}{Bluetooth Low Energy}
\newacronym{BR}{BR}{Basic Rate}
\newacronym{CID}{CID}{Channel ID}
\newacronym{PDU}{PDU}{Protocol Data Unit}
\newacronym{B-Frame}{B-Frame}{Basic Information Frame}
\newacronym{C-Frame}{C-Frame}{Control Frame}
\newacronym{PSM}{PSM}{Protocol/Service Multiplexer}
\newacronym{HCI}{HCI}{Host Controller Interface}
\newacronym{EDR}{EDR}{Enhanced Data Rate}
\newacronym{SCO}{SCO}{Synchronous Connection-Oriented}
\newacronym{HSP}{HSP}{Headset Profile}
\newacronym{ATT}{ATT}{Attribute Protocol}
\newacronym{TCP}{TCP}{Transmission Control Protocol}
\newacronym{SoC}{SoC}{System-on-a-Chip}
\newacronym{IoT}{IoT}{Internet of Things}
\newacronym{GPS}{GPS}{Global Positioning System}
\newacronym{CERT}{CERT}{Computer Emergency Response Team}
\newacronym{API}{API}{Application Programming Interface}
\newacronym{MITM}{MITM}{Machine-in-the-Middle}
\newacronym{TLS}{TLS}{Transport Layer Security}
\newacronym{OTA}{OTA}{Over-the-Air}
\newacronym{XSS}{XSS}{Cross-Site Scripting}
\newacronym{SDK}{SDK}{Software Development Kit}
\newacronym{MQTT}{MQTT}{Message Queuing Telemetry Transport}
\newacronym{CVE}{CVE}{Common Vulnerabilities and Exposures}
\newacronym{ECDH}{ECDH}{Elliptic-curve Diffie–Hellman}
\newacronym{AEAD}{AEAD}{Authenticated Encryption with Associated Data}
\newacronym{AES}{AES}{Advanced Encryption Standard}
\newacronym{TOR}{TOR}{The Onion Router}
\newacronym{GDPR}{GDPR}{EU General Data Protection Regulation}
\newacronym{HSTS}{HSTS}{HTTP Strict Transport Security}
\newacronym{AWS}{AWS}{Amazon Web Services}
\newacronym{LTK}{LTK}{Long Term Key}
\newacronym{LK}{LK}{Link Key}
\newacronym{RCE}{RCE}{Remote Code Execution}
\newacronym{SIV}{SIV}{Synthetic Initialization Vector}
\newacronym{PoC}{PoC}{Proof of Concept}
\newacronym{XPC}{XPC}{Cross-Process Communication}
\newacronym{MFi}{MFi}{Made for iPhone/iPad/iPod}
\newacronym{ACI}{ACI}{Apple Controller Interface}
\newacronym{ECB}{ECB}{Electronic Codebook}
\newacronym{UART}{UART}{Universal Asynchronous Receiver-Transmitter}

\DeclareMathOperator{\kdf}{KDF}
\newacronym{KDF}{KDF}{Key Derivation Function}
\newacronym{PRF}{PRF}{Pseudo-Random Function}

\usepackage{tikz}
\usetikzlibrary{shapes,arrows,calc,positioning,matrix}
\tikzset{
    >=triangle 45
}

\definecolor{darkred}{rgb}{0.831, 0, 0.063}

\usepackage[binary-units]{siunitx}
\usepackage{tabularx}

\definecolor{sorange}{rgb}{0.95, 0.57, 0}
\colorlet{orange}{sorange}

\usepackage{listings}
\lstdefinelanguage{ASM}{
    morekeywords={b, ble, blt, bne, bx, bl, ldr, str, push, pop, mov, add, sub},
    keywordstyle=\color{blue},
    sensitive=false, % keywords are not case-sensitive
    morecomment=[l]{//}, % l is for line comment
    morecomment=[s]{/*}{*/}, % s is for start and end delimiter
    morestring=[b]", % defines that strings are enclosed in double quotes
} %
\lstdefinelanguage{none}{
  identifierstyle=
}

\usepackage{subcaption}

\usepackage{xspace}

\usepackage{bytefield}
\usepackage{xcolor,colortbl}
	
\newcommand{\frida}{F\reflectbox{R}IDA\xspace}

\acmYear{2020}\copyrightyear{2020}
\setcopyright{acmcopyright}
\acmConference[WiSec '20]{13th ACM Conference on Security and Privacy in Wireless and Mobile Networks}{July 8--10, 2020}{Linz (Virtual Event), Austria}
\acmBooktitle{13th ACM Conference on Security and Privacy in Wireless and Mobile Networks (WiSec '20), July 8--10, 2020, Linz (Virtual Event), Austria}
\acmPrice{15.00}
\acmDOI{10.1145/3395351.3399343}
\acmISBN{978-1-4503-8006-5/20/07}

\acmSubmissionID{28}

\begin{document}

%%
%% The "title" command has an optional parameter,
%% allowing the author to define a "short title" to be used in page headers.
\title{MagicPairing: Apple's Take on Securing Bluetooth Peripherals}

%%
%% The "author" command and its associated commands are used to define
%% the authors and their affiliations.
%% Of note is the shared affiliation of the first two authors, and the
%% "authornote" and "authornotemark" commands
%% used to denote shared contribution to the research.

\author{Dennis Heinze}
\affiliation{%
  \institution{Secure Mobile Networking Lab}
  \institution{TU Darmstadt, Germany}
  \streetaddress{Pankratiusstraße 2}
%  \city{Darmstadt}
%  \state{Hesse}
  \postcode{64289}
}
\email{dheinze@seemoo.de}

\author{Jiska Classen}
\affiliation{%
  \institution{Secure Mobile Networking Lab}
  \institution{TU Darmstadt, Germany}
  \streetaddress{Pankratiusstraße 2}
%  \city{Darmstadt}
%  \state{Hesse}
  \postcode{64289}
}
\email{jclassen@seemoo.de}

\author{Felix Rohrbach}
\affiliation{%
  \institution{Cryptoplexity}
  \institution{TU Darmstadt, Germany}
  \streetaddress{Pankratiusstraße 2}
%  \city{Darmstadt}
%  \state{Hesse}
  \postcode{64289}
}
\email{felix.rohrbach@cryptoplexity.de}

%\author{Matthias Hollick}
%\affiliation{%
%  \institution{Secure Mobile Networking Lab, TU Darmstadt}
%  \streetaddress{Mornewegstrasse 32}
%  \city{Darmstadt}
%  \state{Hesse}
%  \postcode{64293}
%}

%%
%% By default, the full list of authors will be used in the page
%% headers. Often, this list is too long, and will overlap
%% other information printed in the page headers. This command allows
%% the author to define a more concise list
%% of authors' names for this purpose.
\renewcommand{\shortauthors}{Heinze et al.}

%%
%% The abstract is a short summary of the work to be presented in the
%% article.
%!TEX root = ../magic.tex

% The abstract is a short summary of the work to be presented in the article.
\begin{abstract}
%what is the problem? -> device pairing is just so uncomfortable!
%why is it a problem? -> also look how the big tech companies are solving this
%how did we solve/analyze it?
%what was the outcome?

Device pairing in large \ac{IoT} deployments is a challenge for device manufacturers and users. Bluetooth offers a comparably smooth trust on first use pairing experience. Bluetooth, though, is well-known for security flaws in the pairing process. \newline 
In this paper, we analyze how \emph{Apple} improves the security of Bluetooth pairing while still maintaining its usability and specification compliance. The proprietary protocol that resides on top of Bluetooth is called \emph{MagicPairing}. It enables the user to pair a device once with \emph{Apple's} ecosystem and then seamlessly use it with all their other \emph{Apple} devices. \newline
We analyze both, the security properties provided by this protocol, as well as its implementations. In general, \emph{MagicPairing} could be adapted by other \ac{IoT} vendors to improve Bluetooth security. Even though the overall protocol is well-designed, we identified multiple vulnerabilities within \emph{Apple's} implementations with over-the-air and in-process fuzzing.

\end{abstract}

%%
%% The code below is generated by the tool at http://dl.acm.org/ccs.cfm.
%% Please copy and paste the code instead of the example below.
%%
\begin{CCSXML}
<ccs2012>
<concept>
<concept_id>10002978.10003006</concept_id>
<concept_desc>Security and privacy~Systems security</concept_desc>
<concept_significance>500</concept_significance>
</concept>
<concept>
<concept_id>10002978.10003022.10003023</concept_id>
<concept_desc>Security and privacy~Software security engineering</concept_desc>
<concept_significance>300</concept_significance>
</concept>
<concept>
<concept_id>10002978.10003022.10003465</concept_id>
<concept_desc>Security and privacy~Software reverse engineering</concept_desc>
<concept_significance>100</concept_significance>
</concept>
<concept>
<concept_id>10003033.10003039.10003051</concept_id>
<concept_desc>Networks~Application layer protocols</concept_desc>
<concept_significance>500</concept_significance>
</concept>
</ccs2012>
\end{CCSXML}

\ccsdesc[500]{Security and privacy~Systems security}
\ccsdesc[300]{Security and privacy~Software security engineering}
\ccsdesc[100]{Security and privacy~Software reverse engineering}
\ccsdesc[500]{Networks~Application layer protocols}

%%
%% Keywords. The author(s) should pick words that accurately describe
%% the work being presented. Separate the keywords with commas.
\keywords{Bluetooth, Pairing, Security}

%%
%% This command processes the author and affiliation and title
%% information and builds the first part of the formatted document.
\maketitle

%u%!TEX root = ../magic.tex

\section{Introduction}

%\todo{motivation:\\
%- tofu is insecure against knob/ecdh\\
%- permanent link key - can get stolen by rce\\
%- pair your gadget with all your devices individually\\
%- can we do better while staying specification compliant with the chip?\\
%- magic pairing solves this!\\
%- we also tested the security for logical bugs.\\
%- after first logical bugs and also null-pointers we switched to on-device handler fuzzing with FRIDA. everything a crappy implementation deserves.}

Bluetooth device pairing has a long history of security flaws~\cite{confusion, knob, 2018:biham, hypponen2007nino, 2005:shaked, ryan2013bluetooth, antonioli20bias}.
While most issues were fixed in the Bluetooth 5.2 specification~\cite{bt52}, it is reasonable to assume that even this version
is not bullet-proof. Adding further layers of encryption within the applications using Bluetooth is one solution many \ac{IoT} developers
chose~\cite{classen2018anatomy}---but this leads to their devices being incompatible in communicating with third-party applications and drains battery.
Thus, encrypting data twice is no satisfying solution to this problem.

Looking back into the history of Bluetooth security issues, it is not the encryption
itself that has been exploited this frequently. Most problems originated from the initial key negotiation and connection setup.
In Bluetooth, trust is established on first use by generating a permanent key.
This permanent key protects device authenticity, message integrity, and message confidentiality~\cite[p. 269]{bt52}.
It is established individually between each pair of devices and only changes when a user manually deletes and reestablishes a pairing.
In Classic Bluetooth, the permanent key is called \ac{LK}, while it is called \ac{LTK} in \ac{BLE}---however, they can be converted into each other~\cite[p. 280]{bt52}.
For the duration of each Bluetooth connection, a session key is derived from the permanent key.
Thus, if a device is out of reach or switched off, this invalidates a session key.

%\todo{oli says there are network keys for bluetooth but i've never seen that being supported in practice. also, we're not talking about a concurrently interconnected mesh or piconet... so i think it's not relevant here.}

In modern \ac{IoT} deployments, Bluetooth device pairing has two major shortcomings: \emph{(1)} It does not scale for pairing to many devices within an existing infrastructure, and \emph{(2)} once the permanent key is leaked, all security assumptions break for past and future connections.
The permanent key can either be attacked by an active \ac{MITM} during  pairing~\cite{2018:biham,hypponen2007nino,ryan2013bluetooth} or by \ac{RCE} vulnerabilities within the chip~\cite{classen201936c3}.

\emph{Apple} solves both challenges by introducing a protocol called \emph{MagicPairing}. It pairs \emph{AirPods} once and then enables the user to instantly use them on all their \emph{Apple} devices. Security is improved by generating fresh ``permanent'' keys based on the user-specific \emph{iCloud} keys for each session. Seamless ecosystem integration and security are imperative, since \emph{AirPods} are able to interact with the \emph{Siri} assistant.

Despite being a proprietary extension, \emph{MagicPairing} is specifica\-tion-compliant to the \ac{HCI}, and thus, can use off-the-shelf Bluetooth chips. The general logic of \emph{MagicPairing} could be integrated into any cloud-based \ac{IoT} ecosystem, increasing relevance for the security community in general. Our contributions on research of \emph{MagicPairing} are as follows:

\begin{itemize}
\item We reverse-engineer the \emph{MagicPairing} protocol.
\item We analyze the security aspects provided by this protocol and their applicability to other wireless ecosystems.
\item We document the proprietary \emph{iOS}, \emph{macOS}, and \emph{RTKit} Bluetooth stacks.
\item We manually test \emph{MagicPairing} for logical bugs and automatically fuzz its three  implementations.
\item We responsibly disclosed multiple vulnerabilities.
\end{itemize}

While the overall idea of \emph{MagicPairing} is new and solves shortcomings of the Bluetooth specification, we found various issues in \emph{Apple's} implementations.
As \emph{MagicPairing} is available prior to pairing and encryption, it poses a large zero-click wireless attack surface.
We found that all implementations have different issues, including a lockout attack and a \ac{DoS} causing \SI{100}{\percent} CPU load. We identified these issues performing both, generic over-the-air testing and \emph{iOS} in-process fuzzing.

Our fuzzing techniques can also be used to test other Bluetooth stacks and protocols. The \acp{PoC} for the identified vulnerabilities as well as the over-the-air fuzzing additions are available within the \emph{InternalBlue} project on \emph{GitHub}.
The \emph{ToothPicker} in-process fuzzer part that integrates into \emph{InternalBlue} will follow soon, but has to be slightly delayed due to further findings~\citep{ttdennis}.

This paper is structured as follows. \autoref{sec:protocol} gives an overview of the reverse-engineered \emph{MagicPairing} protocol.
Its security properties are explained in \autoref{sec:crypto}.
Implementation internals regarding \emph{Apple's} Bluetooth stacks as well as \emph{MagicPairing}-specific details are provided in \autoref{sec:implementation}.
Then, we explain our fuzzing setup in \autoref{sec:fuzzing} used to identify the vulnerabilities explained in \autoref{sec:vulns}.
We conclude our work in \autoref{sec:conclusion}.
%!TEX root = ../magic.tex

\begin{figure*}[!t]

\center

	\begin{center}
	\begin{tikzpicture}[minimum height=0.55cm, scale=0.8, every node/.style={scale=0.8}, node distance=0.7cm]
	
	% Devices 
    \node[inner sep=0pt] (icloud) at (-1,2.4)
    {\includegraphics[height=1.1cm]{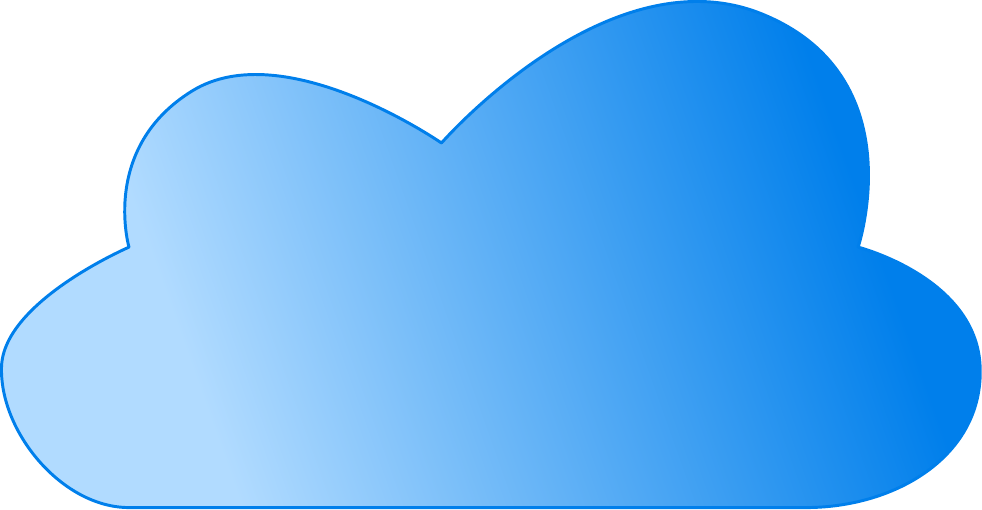}};
    \node[inner sep=0pt] (iphone) at (3,1)
    {\includegraphics[height=1.5cm]{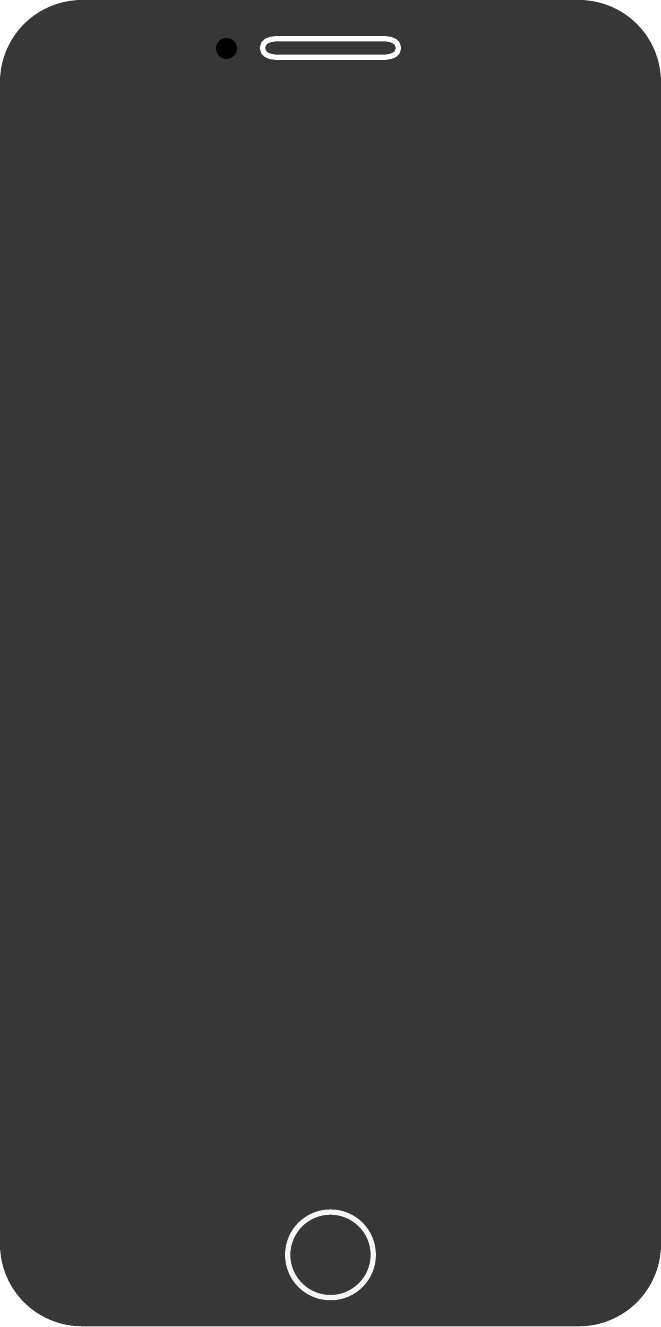}};  
    \node[inner sep=0pt] (iphonex) at (3,1)
    {\includegraphics[height=1.2cm]{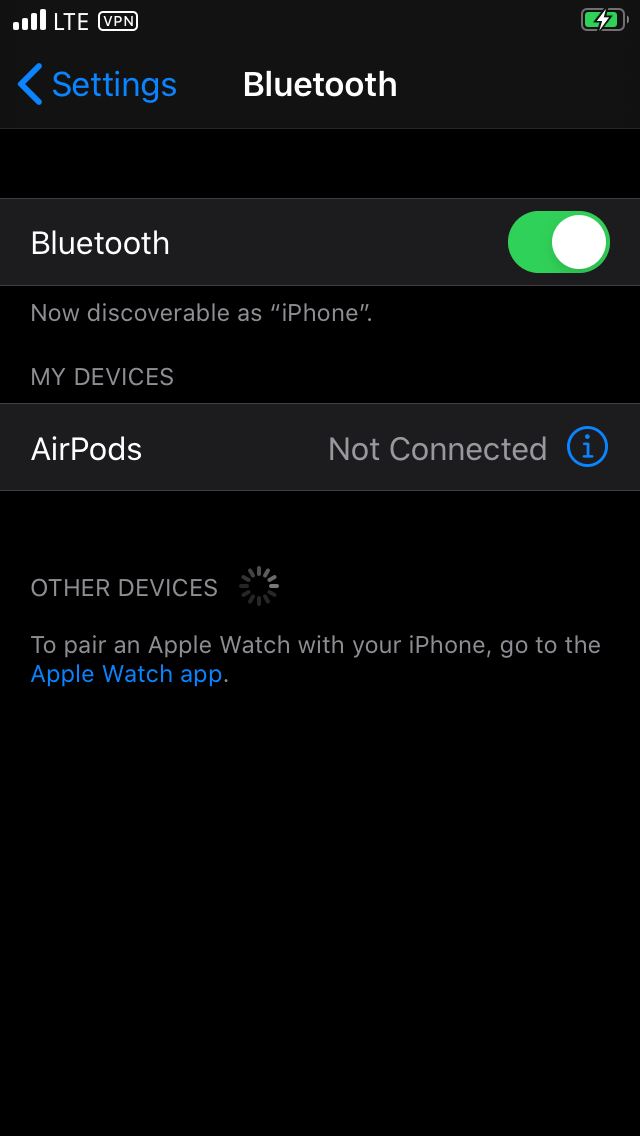}};   
    \node[inner sep=0pt] (airpod) at (13,1)
    {\includegraphics[height=1.5cm]{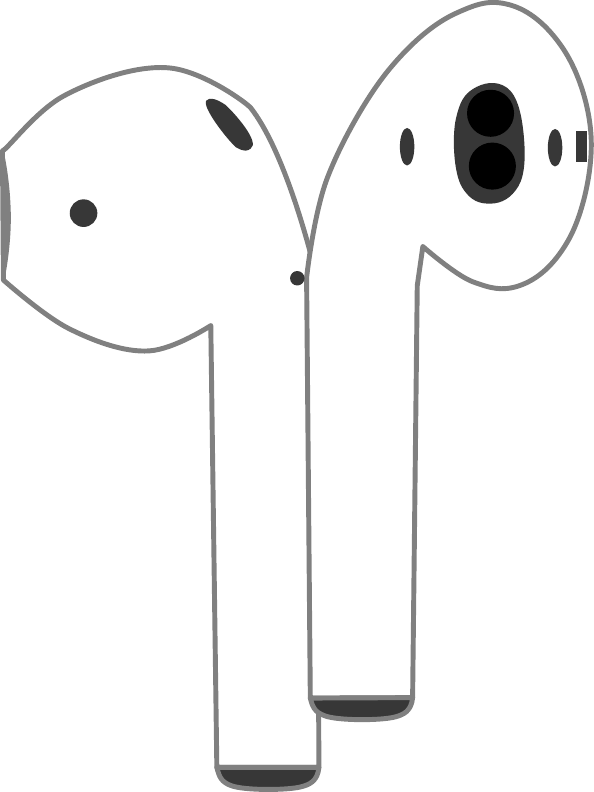}};

    % Device Labels
    \node[below=of iphone.south, anchor=south, yshift=0.3cm] (iphonetxt) {iPhone};
    \node[below=of icloud.south, anchor=south, yshift=0.3cm] (icloudtxt) {iCoud};
    \node[below=of airpod.south, anchor=south, yshift=0.3cm] (airpodtxt) {AirPods};

    % Device Lines
    \draw[-,gray,dashed,thick] (3,-0.4) -- (3,-10);
    \draw[-,gray,dashed,thick] (13,-0.4) -- (13,-10);
    
    % MasterKey
    \path[<-,black] (-1,1.3) edge [bend right] node [left,align=left] {\textcolor{gray}{\texttt{MasterKey}}\\ \textcolor{gray}{\texttt{MasterHint}}\vspace{-1em}} (2.4,-0.1);
    \node[align=left] at (1.1,1) {1a Key Creation};

    \path[->] (3.2,-0.5) edge node[sloped, anchor=center, above, text width=8.5cm,yshift=-1.4em] {1b Key Distribution\\ \textcolor{gray}{\texttt{accKey} via SSP and AAP}} (12.8,-0.5);
    
    % textcircled fix - we need ellipse for 1a 1b
    \draw (0.2,1) ellipse (0.7em and 0.5em);
    \draw (3.9,-0.3) ellipse (0.7em and 0.5em);

    % divide setup
    \draw[-,black,dotted,thick] (-2,-1.2) -- (19.2,-1.2);
    \node[align=right,right] (setup) at (-2, -0.8) {\textbf{Setup}\hspace*{3.1em}\footnotesize{???}};
    \node[align=right,right] (pairing) at (-2, -1.7) {\textbf{Pairing}};
    \node[inner sep=0pt,anchor=west] (uc) at (-0.5, -0.8)
    {\includegraphics[height=1em]{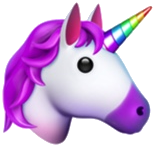}}; 
    \node[inner sep=0pt,anchor=west] (mp) at (-0.5, -1.7)
    {\includegraphics[height=1em]{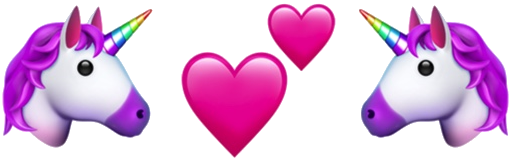}};

    \node[gray,align=left,left] at (2.8, -2) {\texttt{BdAddrBlob}};
    
    \path[->] (3.2,-2) edge node[sloped, anchor=center, above, text width=8.5cm,yshift=-3.8em] {\textcircled{2} Hint\\ \textcolor{gray}{\texttt{hint$_{0x10}$=enc$_{ECB}$(MasterHint, BdAddrBlob)}}  \\ \textcolor{gray}{\texttt{nonce\_host$_{0x10}$=arc4\_rand()}}  \\ \textcolor{gray}{\texttt{ratchet\_host$_{0x4}$}}} (12.8,-2);

    \node[gray,align=left,right] at (13.2, -3.5) {\texttt{while local\_ratchet < ratchet:} \\ \hspace*{2em} \texttt{accKey=enc$_{ECB}$(accKey, 0x10*0x00)} \\ \\ \\ Generate authentication and encryption key: \\ \texttt{SIV\_Key$_{0x20}$=enc$_{ECB}$(accKey,} \\ \hspace*{1em} \texttt{''bt\_aessivauthentbt\_aessivencrypt'')} };

    \path[<-] (3.2,-4.5) edge node[sloped, anchor=center, above, text width=8.5cm,yshift=-3.8em] {\textcircled{3} Ratcheting\\ \textcolor{gray}{\texttt{AES\_SIV$_{0x20}$=enc$_{SIV}$(SIV\_KEY, rand\_airpod+nonce\_host }}\\ \hspace*{9em}\textcolor{gray}{\texttt{+addr\_airpod)}}\\ \textcolor{gray}{\texttt{ratchet\_airpod$_{0x4}$}}} (12.8,-4.5);
  
    \path[->] (3.2,-6.5) edge node[sloped, anchor=center, above, text width=8.5cm,yshift=-2.6em] {\textcircled{4} AES-SIV\\ \textcolor{gray}{\texttt{AES\_SIV$_{0x20}$=enc$_{SIV}$(SIV\_KEY, nonce\_host+rand\_host  }} \\ \hspace*{9em}\textcolor{gray}{\texttt{+rand\_airpod+hint)}}} (12.8,-6.5);
    
   \node[gray,align=left,right] at (13.2, -7) {unpack \& check };
    
    \path[<-] (3.2,-8) edge node[sloped, anchor=center, above, text width=8.5cm,yshift=-1.2em] {\textcircled{5} Status Success \& Link Key Derivation\\  \textcolor{gray}{\texttt{status}} } (12.8,-8);

   %\node[gray,align=left,right] at (13.2, -9) {\textcolor{gray}{\texttt{session\_key\_pre1=}} \\ \hspace*{1em}\textcolor{gray}{\texttt{enc$_{ECB}$(rand\_host, rand\_airpod) }} \\ \textcolor{gray}{\texttt{session\_key\_pre2=}} \\ \hspace*{1em}\textcolor{gray}{\texttt{enc$_{ECB}$(rand\_airpod, '$\textbackslash$x00'*0x10) }} \\\textcolor{gray}{\texttt{while i != 16: }} \\ \hspace*{1em}\textcolor{gray}{\texttt{link\_key[i]=session\_key\_pre1[i]+session\_key\_pre2[i] }}};    	
    	
   \node[gray,align=left,right] at (13.2, -9.1) {\textcolor{gray}{\texttt{sk\_pre1=enc$_{ECB}$(rand\_host, rand\_airpod) }} \\ \textcolor{gray}{\texttt{sk\_pre2=enc$_{ECB}$(rand\_airpod, '$\textbackslash$x00'*0x10) }} \\\textcolor{gray}{\texttt{while i != 16: }} \\ \hspace*{1em}\textcolor{gray}{\texttt{link\_key[i]=sk\_pre1[i]+sk\_pre2[i] }}};
   
    \node[gray,align=left,left] at (2.8, -5) {calculate \texttt{accKey} and \texttt{SIV\_Key}};
    
    \node[gray,align=right,left, text width=5cm] at (2.8, -9.7) {AAP: \texttt{UpdateMagicCloudKeys} sends the new \texttt{accKey} to the AirPods};

	 \filldraw[align=left, fill=gray, gray](13.2,-2) rectangle node (host) {\textcolor{white}{$KDF_1$}} ++(1,0.5);
	 
	 \filldraw[align=left, fill=gray, gray](13.2,-3.6) rectangle node (host) {\textcolor{white}{$KDF_2$}} ++(1,0.5);

	 \filldraw[align=left, fill=gray, gray](13.2,-8.2) rectangle node (host) {\textcolor{white}{$KDF_3$}} ++(1,0.5);
	 
	 \filldraw[align=left, fill=gray, gray](1.8,-8.2) rectangle node (host) {\textcolor{white}{$KDF_3$}} ++(1,0.5);
    
	\end{tikzpicture}
	\end{center}
	
\caption{\emph{MagicPairing} protocol steps.}
\label{fig:magicpairing}
\end{figure*}
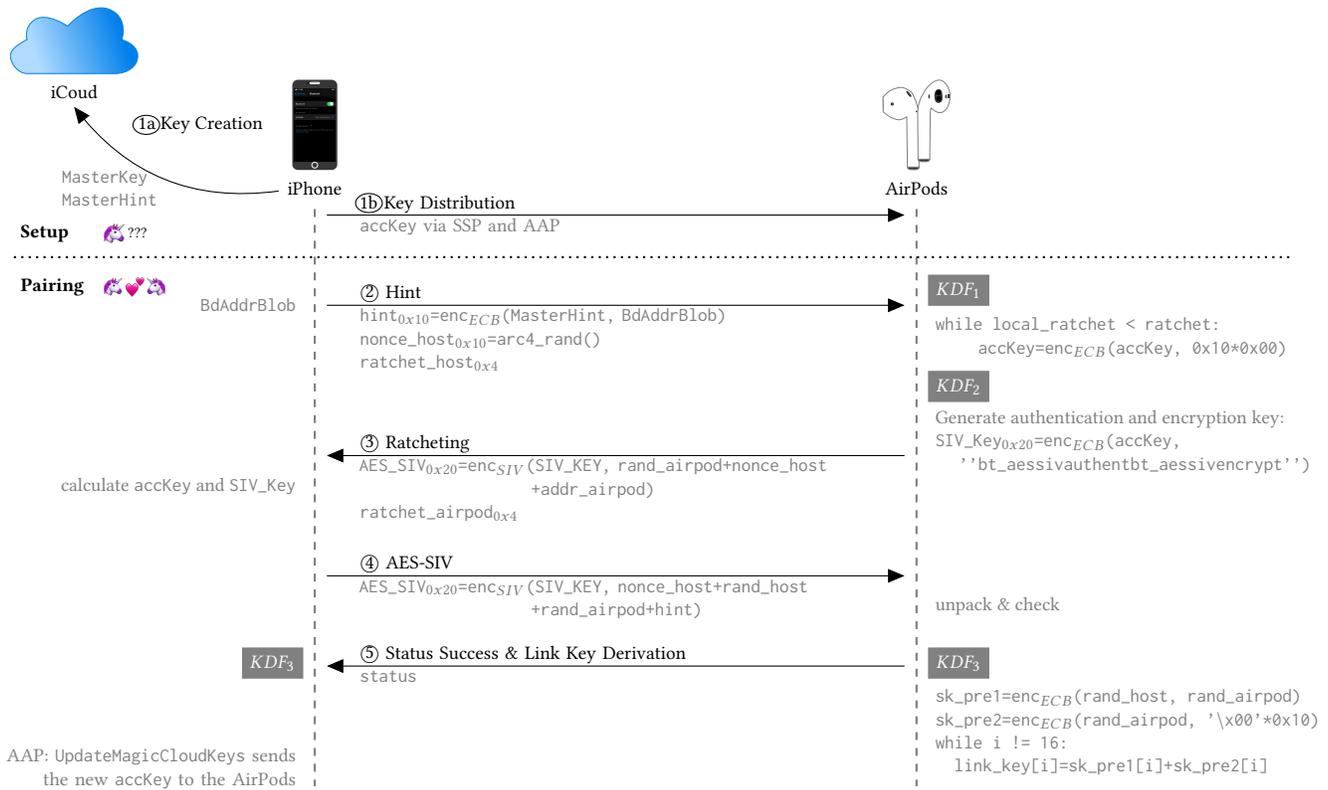

\section{The MagicPairing Protocol}
\label{sec:protocol}

\emph{MagicPairing} is a proprietary protocol providing seamless pairing capabilities, for instance between a user's \emph{AirPods} and all their \emph{Apple} devices. This is achieved by synchronizing keys over \emph{Apple's} cloud service \emph{iCloud}. The
ultimate goal of the \emph{MagicPairing} protocol is to derive a Bluetooth \acf{LK} that
is used between a single device and the \emph{AirPods}.
A fresh \ac{LK} is created for each connection, which significantly reduces the lifetime of this \ac{LK}.

When a new or reset pair of \emph{AirPods} is initially paired with an \emph{Apple} device belonging to
an \emph{iCloud} account, \ac{SSP} is used~\cite[p. 271ff]{bt52}. All subsequent connections between the \emph{AirPods} and
devices connected to that \emph{iCloud} account  will use the \emph{MagicPairing} protocol as
pairing mechanism. \emph{MagicPairing} involves multiple keys and derivation functions. It
relies on \ac{AES} in \ac{SIV} mode for authenticated encryption~\cite{rfc5297}.

The protocol mainly consists of five phases. 
The protocol flow is
visualized in \autoref{fig:magicpairing} and explained in the following.
\emph{MagicPairing} depends on a shared secret between the two participants.
Therefore, the first phase establishes and exchanges a secret, followed by phases of the actual protocol.
% The next phases then
%determine steps in the protocol itself.
As the protocol is not publicly documented, our naming
relies on debug output and strings found in the respective components, i.e., the
Bluetooth daemon \texttt{bluetoothd} for \emph{iOS} and \emph{macOS}, as well as the \emph{AirPod} firmware. Further implementation and Bluetooth stack details follow later in \autoref{sec:implementation}.

\subsection{Phase 1: Key Creation and Distribution}
\emph{MagicPairing} relies on a shared secret between the \emph{AirPods} and a user's
\emph{iCloud} devices, the \emph{Accessory Key} (also \texttt{accKey}). This key is created by the first device pairing  \emph{AirPods} for a specific \emph{iCloud} account. After establishing
an encrypted Bluetooth connection using \gls{SSP}, the \emph{Accessory Key} needs to be
transmitted to the \emph{AirPods}. \emph{Apple} is using the
\emph{AAP Protocol}\footnote{\emph{AAP} is used for communication between a device and \emph{AirPods}. Its services all revolve around configuring \emph{AirPods} and obtaining information from them, such as firmware updates, getting and setting tapping actions, or exchanging key material.} for the \emph{Accessory Key} transfer.
In addition to the
\emph{Accessory Key}, the host also creates an \emph{Accessory Hint}, which uniquely
identifies the connection between an \emph{iCloud} account and the target
device. The initiating
device uses the \emph{iCloud} account's \emph{Master Key} and \emph{Master Hint} to create the \emph{Accessory Key} and the \emph{Accessory Hint}.
In case these \emph{Master} credentials do not exist yet, the device provisions them by creating
random bytes. Another component that is needed to create the \emph{Accessory Key} and
\emph{Accessory Hint} is the so-called \emph{Bluetooth Address Blob}, which is a
deterministic mutation of the Bluetooth address of the targeted device, as shown in \autoref{lst:bdaddrBlob}.
The \emph{Bluetooth Address Blob} is then encrypted with
the \emph{Master Key} using AES in ECB mode to create the \emph{Accessory Key}. The \emph{Accessory Hint} is created by
encrypting the \emph{Bluetooth Address Blob} with the \emph{Master Hint}, respectively.

After the initial setup, both devices share the same \emph{Accessory Key}. All devices logged into the \emph{iCloud} account can generate the same \emph{Accessory Key}.
%From now on, any device and the \emph{AirPods} can initiate a pairing.
In the following example, the device connects to the \emph{AirPods}, but all steps could also happen in the opposite direction.

\begin{lstlisting}[language=C, caption=Creating a \emph{Bluetooth Address Blob.}, label=lst:bdaddrBlob]
blob[1:5] = address[5:0]
blob[6:9] = address[1:4] ^ address[0:3]
\end{lstlisting}

\subsection{Phase 2: Hint}

The first repeating phase in the \emph{MagicPairing} protocol is the \emph{Hint} phase.
It ensures that both sides will agree on the same fresh session key in the end that belongs
to the correct device.
The device initiates the pairing by sending a \emph{Hint} message. The \emph{Hint} message includes three entries, the \texttt{hint}, a random nonce
generated by the initiating host, and a \emph{Ratchet}.
The \emph{Ratchet} is a counter used in
later steps of the pairing process to rotate keys.

The receiving end
performs a local \emph{Accessory Key} table lookup for the connecting device. The \emph{AirPods}
use the \texttt{hint} that is included in the \emph{Hint} message as a reference, \emph{iOS} and
\emph{macOS} devices use the connecting device's Bluetooth address to look up the key.
If no key is found, the protocol is aborted with a \emph{Status Message} 
indicating that the initiating device is unknown.

\subsection{Phase 3: Ratcheting}
The \emph{Ratcheting} phase is essentially a key rotation and derivation phase.
The goal of \emph{Ratcheting} is to renew and maintain short-lived session keys~\cite{doubleratchet}.
 First, the
\emph{Accessory Key} is rotated and then a \emph{SIV Key} is derived from the rotated key.
The \emph{Accessory Key} is rotated by encrypting a buffer of \num{16} null-bytes with the
current \emph{Accessory Key} using \ac{AES} in \ac{ECB} mode. After one rotation step, the current counter,
or \emph{Ratchet}, is incremented. This is done until the local \emph{Ratchet} equals the \emph{Hint's Ratchet}. Then,
the \emph{SIV Key} is derived from the \emph{Accessory Key}
by encrypting the static \SI{32}{\byte} string \path{bt_aessivauthentbt_aessivencrypt} with the
\emph{Accessory Key} using AES in ECB mode. Next, an
\emph{AES-SIV} value is created. For this, the device creates a local random value,
concatenates it with the received nonce and its own Bluetooth address, and encrypts it
with the \emph{SIV Key}. This time, AES is used in SIV mode without a nonce or any
additional data.
At the end of this phase, a \emph{Ratchet AES-SIV} message is sent back to the initiating
device. It contains the local \emph{Ratchet} value, as well as the \emph{AES-SIV} value.
The initiating device executes the same key derivation steps
as mentioned above using the received \emph{Ratchet} value. This leads to both devices
having the same updated \emph{Accessory Key} and \emph{SIV Key}. Using the derived \emph{SIV Key},
the initiating device can now decrypt the \emph{AES-SIV} value to unpack the random value
of the responding device.

\subsection{Phase 4: AES-SIV}
The initiating device will now create another \emph{AES-SIV} value. However, this one is
different from the one that the responding device created before. First, the device creates a new random value. Then it concatenates its nonce value, the new random value, the previously received \emph{AirPods} random value, and the \texttt{hint}
value. This \SI{64}{\byte} value is then encrypted with the derived \emph{SIV Key} using AES in
SIV mode and sent to the responding device.
If the \emph{AirPods} can decrypt the received data, they send a
\emph{MagicPairing Status} success message.

\subsection{Phase 5: Link Key Derivation}
Finally, a Bluetooth-compliant connection \ac{LK} is
derived. First, two \emph{Session Pre Keys} are created and XORed. The
\emph{Session Pre Key 1} is created by encrypting the responding device's random value
with the initiating device's random value as key using AES. The \emph{Session Pre
Key 2} is created by encrypting a \SI{16}{\byte} null-byte buffer with the responding device's
random value using AES.

It is important to note that even though \emph{MagicPairing} is a custom key derivation protocol, the further usage of this key is still compliant to the Bluetooth specification and does not
require any modifications to the Bluetooth chip.
When establishing an encrypted connection, the chip sends an \gls{HCI} command to ask the host for the stored \ac{LK}~\cite[p. 1948]{bt52}. In case of \emph{MagicPairing}, the \ac{LK} is not
taken from the host's storage but freshly created, which is
completely transparent to the chip.
In either case, the \acp{LK} is stored on the host.
However, it is only short-lived within \emph{MagicPairing}, while it is permanent for a normal Bluetooth pairing.

\begin{figure*}[!b]

\center

	\begin{center}
	\begin{tikzpicture}[minimum height=0.55cm, scale=0.8, every node/.style={scale=0.8}, node distance=0.7cm]

    % divide space
    \draw[-,black,dotted,thick] (-9.5,2.25) -- (6,2.25);
    \node[align=right,right] (top) at (-9.5, 2.6) {\textbf{User-Space}};
    \node[align=right,right] (down) at (-9.5, 1.9) {\textbf{Kernel-Space}};
    
    % gray background lines
    
    \draw[-,gray,dotted] (-4.5,0) -- (-4.5,4.75);
    \draw[-,gray,dotted] (2.5,0) -- (2.5,4.75);
    \draw[-,gray,dotted] (9.5,0) -- (9.5,4.75);

	 \filldraw[align=left, fill=blue!30](-7,0) rectangle node (host) {Broadcom Bluetooth Chip} ++(5,0.5);
	 \filldraw[align=left, fill=white](-7,0.75) rectangle node (host) {XNU} ++(5,0.5);
	 \filldraw[align=left, fill=white](-7,1.5) rectangle node (host) {\texttt{IOBluetoothFamily}} ++(5,0.5);
	 \filldraw[align=left, fill=white](-7,2.5) rectangle node (host) {\texttt{IOKit}} ++(5,0.5);
	 \filldraw[align=left, fill=white](-7,3.25) rectangle node (host) {\texttt{IOBluetooth}} ++(5,0.5);
	 \filldraw[align=left, fill=white](-7,4.0) rectangle node (host) {\texttt{bluetoothd}} ++(5,0.5);
	 \filldraw[align=left, fill=white](-7,4.75) rectangle node (host) {\texttt{CoreBluetooth}} ++(5,0.5);
	 
	     \node[align=right,right] (ne) at (-1.25,5) {\textcolor{darkred}{$\neq$}};

	     \node[align=right,right] (ne) at (-1.25,4.25) {\textcolor{darkred}{$\neq$}};

	 \filldraw[align=left, fill=blue!30](0,0) rectangle node (host) {Broadcom Bluetooth Chip} ++(5,0.5);
	 \filldraw[align=left, fill=white](0,0.75) rectangle node (host) {XNU} ++(5,0.5);
	 \filldraw[align=left, fill=white](0,4.0) rectangle node (host) {\texttt{bluetoothd}} ++(5,0.5);
	 \filldraw[align=left, fill=white](0,4.75) rectangle node (host) {\texttt{CoreBluetooth}} ++(5,0.5);

	 \filldraw[align=left, fill=blue!30](7,0) rectangle node (host) {\\ \\ \\ \\{Marconi Bluetooth Chip}} ++(5,2);
	 \filldraw[align=left, fill=white](7.5,0.75) rectangle node (host) {PHY Firmware} ++(4,0.5);
	 \filldraw[align=left, fill=white](7.5,1.25) rectangle node (host) {Controller Firmware} ++(4,0.5);
	 \filldraw[align=left, fill=white](7,4.75) rectangle node (host) {RTKit OS} ++(5,0.5);

    \node[inner sep=0pt] (iphone) at (0,0)
    {\includegraphics[height=1.5cm]{pics/iphone.pdf}};  
    \node[inner sep=0pt] (iphonex) at (0,0)
    {\includegraphics[height=1.2cm]{pics/iosbt.png}};  
    
    \node[inner sep=0pt] (airpod) at (7,0)
    {\includegraphics[height=1.5cm]{pics/airpod.pdf}}; 
     
    \node[inner sep=0pt] (airpod) at (-7.5,0)
    {\includegraphics[height=1.5cm]{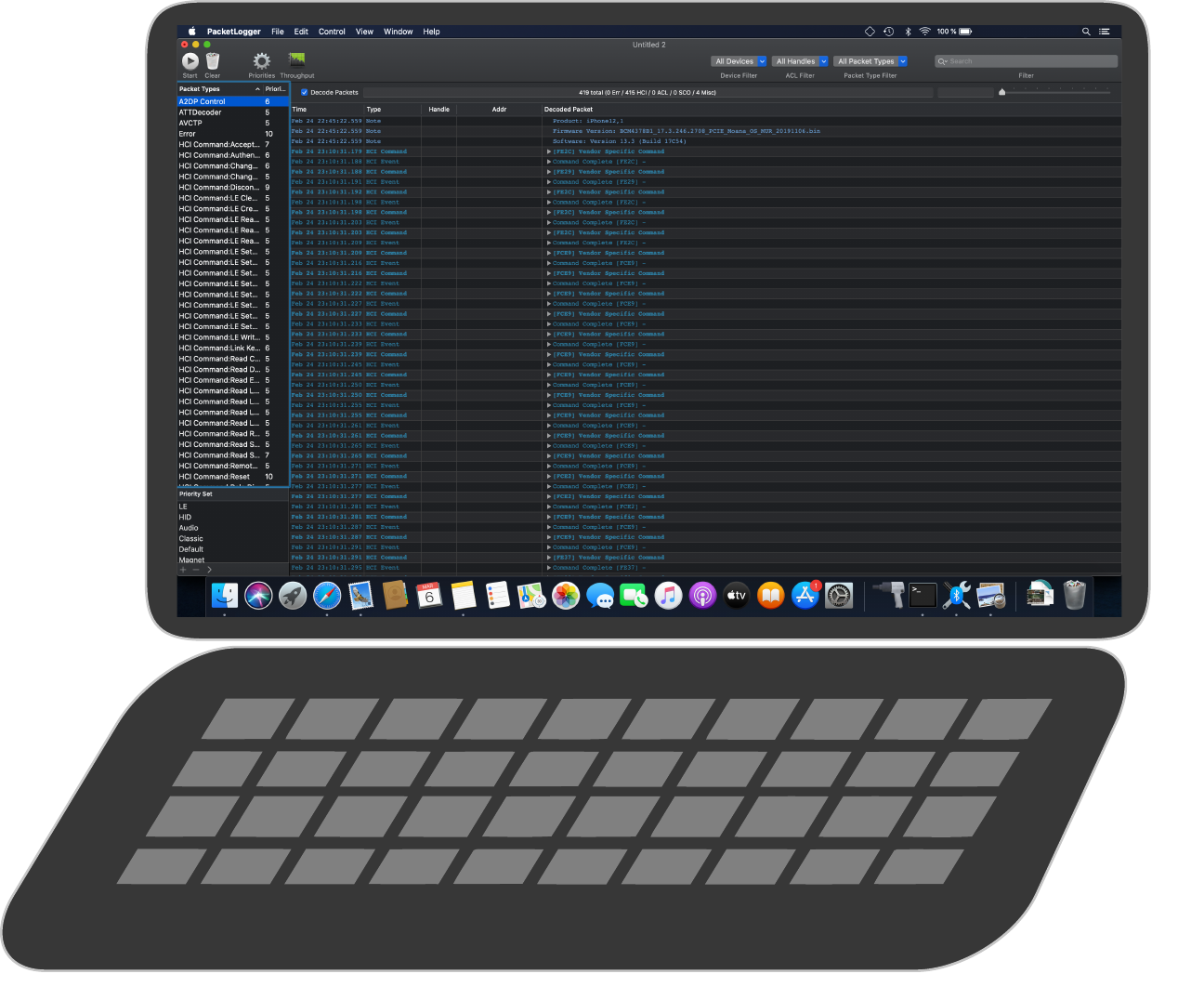}};

	\end{tikzpicture}
	\end{center}
	
\caption{\emph{Apple's} Bluetooth stacks: \emph{macOS}, \emph{iOS}, and \emph{RTKit}.}
\label{fig:stacks}
\end{figure*}

%!TEX root = ../magic.tex
\section{Security Properties}
\label{sec:crypto}

The security goals of the \emph{MagicPairing} protocol seem to be to provide authentication and a fresh shared key for each connection. It uses a symmetric ratcheting algorithm and authenticated encryption to achieve these goals.

The idea of ratcheting was introduced by Borisov, Goldberg, and Brewer~\cite{DBLP:conf/wpes/BorisovGB04}. They introduced a continuous Diffie-Hellman key exchange providing forward and post-compromise secrecy within a session. Marlinspike and Perrin~\cite{doubleratchet} extended this notion in the \emph{Double Ratchet} algorithm to include a second, symmetric ratchet that updates the key while one party is offline. A \emph{Double Ratchet} only provides forward secrecy, but no post-compromise secrecy.

The \emph{MagicPairing} protocol uses only the symmetric ratcheting and therefore does not provide post-compromise secrecy. However, the usage of no expensive public-key cryptography makes this protocol feasible for usage with \ac{IoT} devices like the \emph{AirPods}. Further, note that \emph{MagicPairing} uses ratcheting in a slightly different way than the previous work: Instead of creating a new key per message, the protocol creates a new key per Bluetooth connection. What is defined in the \emph{Double Ratchet} algorithm as message key is therefore a connection key in this protocol.

In the \emph{Double Ratchet} algorithm, the symmetric ratchet consists of a \ac{KDF} that, given a chain key, produces a new chain key and an independent message key. This is done for each new message, so each new message gets encrypted with a new key. As the \ac{KDF} cannot be inverted, the knowledge of the chain key at some point only allows to calculate future chain and message keys, but no previous chain and message keys. Further, the knowledge of a message key does not enable an adversary to calculate any of the chain keys. \emph{MagicPairing} uses two separate \ac{KDF}s to accomplish the same goal: The first \ac{KDF},
\[\kdf_1(k) = \text{enc}_{ECB}(k, 0^{16}),\]
is used to update the chain key. By using plain AES keyed with the old chain key to encrypt a constant (here: the bit string consisting of only zeros), it uses the \ac{PRF} property of AES, which guarantees that without knowledge of the old chain key $k$, the new chain key is indistinguishable from a random key. The second \ac{KDF},
\[\kdf_2(k) = (\text{enc}_{ECB}(k, c_1), \text{enc}_{ECB}(k, c_2)),\]
where $c_1$ is the string \path{bt_aessivauthent} and $c_2$ \path{bt_aessivencrypt}, produces a connection key, which itself consists of two different key parts, an authentication and an encryption part. By the same argument as for $\kdf_1$, the chain key cannot be calculated from the produced key and both key parts are independent.

The ratchet is initialized with the account key and the position in the ratchet is synchronized by the values \path{ratchet_host} and \path{ratchet_airpod}.

For the encryption of the messages between the host and the \emph{AirPod}, AES is used in the SIV mode of operation. SIV, an authenticated encryption mode, was introduced by Rogaway and Shrimpton~\cite{DBLP:conf/eurocrypt/RogawayS06} and standardized in the combination with AES in RFC5297~\cite{rfc5297}. It is used without any headers in \emph{MagicPairing}, which is secure as long as the entropy of each message is high enough. As all messages encrypted with AES-SIV contain a new random number, the entropy is sufficient.

Finally, \emph{MagicPairing} uses a third \ac{KDF} to generate the key used for the Bluetooth connection, based on two random values, one generated by the host and one generated by the \emph{AirPod}:
\[\kdf_3(r_h, r_a)= \text{enc}_{ECB}(r_h, r_a) \oplus \text{enc}_{ECB}(r_a, 0^{16})\]
Again, this uses the \ac{PRF} property of plain AES to generate a key that is indistinguishable from random as long as not both random values are known.

Knowledge of the final key implies knowledge of the SIV key, which in turn implies the knowledge of the account key, which identifies the party as being connected to the \emph{iCloud} account. Further, for each connection a new key is used in a forward-secret manner. Therefore, the protocol meets the security goals of authentication and forward secrecy.

%!TEX root = ../magic.tex

\section{Implementation Details}
\label{sec:implementation}
In the following, we discuss \emph{Apple}-specific implementation details, which impact our security analysis.
\autoref{ssec:stacks} compares the three Bluetooth stacks.
As all of them differ significantly, the attack surface as well as bugs in their implementations vary.
\autoref{ssec:messages} lists the \emph{MagicPairing} message formats, which are relevant for fuzzing the protocol, as well as understanding the fuzzing results and attacks. \autoref{ssec:advertisement} explains the advertisements sent by \emph{AirPods},
and based on these, connections are initiated. Finally, we spot many spelling mistakes, as shown in \autoref{ssec:spelling}, which outline the \emph{MagicPairing} code quality.

\subsection{Apple's Bluetooth Stacks}
\label{ssec:stacks}

\emph{Apple} uses three fundamentally different Bluetooth stacks in their recent devices. Each stack is for an individual device type and supports a subset of features. Thus, the protocols they support have duplicate implementations. While this circumstance helps us to reverse engineer these protocols, it raises maintenance overhead for \emph{Apple}. From a security perspective, this results in different issues in these stacks, as shown later in \autoref{sec:vulns}.

\emph{RTKit} is a separate framework for resource-constraint embedded devices.
While this separation to reduce features makes sense, also \emph{iOS} and \emph{macOS} have individual Bluetooth stacks. As they are closed-source and there is only little public documentation, we provide an overview in the following. 
\autoref{fig:stacks} compares all stacks.

%\todo{bluetoothd ist über XPC nutzbar und nutzt aber auch nur das high-level IOBluetooth framework. Und davon halt die Methoden die in dem Kasten sind. Die callen dann die \_IOConnectCallStructMethod vom IOKit framework, wo dann via Mach port mit dem Treiber kommuniziert wird (IOBluetoothFamily).}

\subsubsection{macOS}
\label{sssec:macos}
The most recent version of the \emph{macOS} Bluetooth stack was investigated  and documented previously to integrate \emph{InternalBlue}~\cite{dave}.
The \emph{macOS} kernel exposes a user-space \texttt{IOKit} device-interface for Bluetooth~\cite{iokit}. \texttt{IOKit} communicates using a Mach port with the \texttt{IOBluetoothFamily} driver, which supports connectivity to USB, \ac{UART}, and PCIe chips.
User-space applications connect to Bluetooth devices using the \texttt{IOBluetooth} private \ac{API}, which exposes methods to access the chip via \ac{HCI} and send \ac{ACL} data. The \emph{macOS} \texttt{bluetoothd} manages all Bluetooth logic and connects to other daemons such as \texttt{bluetoothaudiod} for music streaming. The public \ac{API} to access Bluetooth on \emph{macOS} is \texttt{CoreBluetooth}, which communicates with \texttt{bluetoothd} via \ac{XPC} and further abstracts the methods exposed by \texttt{IOBluetooth}.

\subsubsection{iOS}
\label{sssec:ios}
\emph{Apple's} mobile operating system is \emph{iOS} and has derivatives called \emph{iPadOS}, \emph{tvOS}, and \emph{watchOS}.
On \emph{iOS}, the Bluetooth chip is exposed as serial character device\footnote{A character device is exposed for all \emph{Broadcom} \ac{UART} chips, which are at least present in the \emph{iPhone 6}, \emph{SE}, \emph{7}, \emph{8}, \emph{X}, \emph{XR}, and various \emph{iPads}. \emph{iOS} also supports \emph{Marconi} (newer \emph{AppleWatches}) and \emph{Broadcom} PCIe (\emph{iPhone XS} and \emph{11}) Bluetooth chips.} to the user-space.
On initialization, \texttt{bluetoothd} directly connects to the exposed Bluetooth socket of this character device. Then, \texttt{bluetoothd} offers Bluetooth-related functionality as an \ac{XPC} service. Similar to \emph{macOS}, this \ac{XPC} service is accessed by the public \texttt{CoreBluetooth} \ac{API}. However, \emph{iOS} \texttt{CoreBluetooth} does not allow apps to create and use Classic Bluetooth connections, which is slightly different from \emph{macOS}. Instead, it offers a higher-level application protocol called \emph{External Accessory} that can be used in combination with \ac{MFi} certified Bluetooth devices~\cite{mfi}.

Even though \ac{HCI} is not openly accessible, it is needed by system components. 
\texttt{bluetoothd} exposes a Mach port for features like \ac{HCI}, which is only accessible by system components. This private framework is called \texttt{MobileBluetooth}.

%It is important to note that while \texttt{bluetoothd} has more of an administrative role on \emph{macOS}, it is key to all Bluetooth-related functionality on \emph{iOS}.

\subsubsection{RTKit and Marconi}
\label{sssec:rtkit}

For embedded devices, \emph{Apple} is using a real-time operating system based on the \emph{RTKit} framework.
\emph{RTKit} is used on multiple embedded controllers and in all recent Bluetooth peripherals, such as the \emph{AirPods 1}, \emph{2}, and \emph{Pro}, \emph{Siri Remote 2}, \emph{Apple Pencil 2}, and \emph{Smart Keyboard Folio}.
While \emph{RTKit} is not well-known, it has an incredibly high market share.
For example, \emph{AirPods} are accounted for \SI{60}{\percent} of the global wireless earbud market~\cite{fortunemarket}.
Moreover, \emph{RTKit} powers a number of other devices and chips in the \emph{Apple} ecosystem, such as the \ac{AOP} firmware included in most of \emph{Apple's} mobile devices like the \emph{iPhone} and \emph{AppleWatch}.

The \emph{RTKit} framework lacks public documentation by \emph{Apple} but has been briefly mentioned by other researchers~\cite{newosxbook}.
The newest \emph{AirPod Pro} firmware strings reveal version information, such as \path{RTKitAudioFrameworkW2}, \path{RTKitOSPlatform-620.60.2 616}, and
\path{RTKit2.2.Internal.sdk}. The latter lets us conclude that \emph{Apple} has an internal \ac{SDK} used to develop \emph{RTKit} applications.

We consolidate all these peripherals into a single Bluetooth stack, however, the their firmware is very different due to their technologies and use cases. The \emph{Siri Remote}, \emph{Apple Pencil}, and \emph{Smart Keyboard} only use \ac{BLE}, while the \emph{AirPods} rely on both \ac{BLE} and Classic Bluetooth. Nonetheless, the basic \emph{RTKit} code is the same.

On the \emph{AirPods}, the communication to the Bluetooth chip is provided via the \ac{ACI} instead of the specification-compliant \acf{HCI}. This is because the \emph{AirPods} use \emph{Apple's} new Bluetooth chip \emph{Marconi}.
An older version of the \emph{PacketLogger} contains a file \path{ACI_HCILib.xml}, which names and partially describes all \ac{ACI} commands. Some of these are \emph{AirPod}-specific, such as synchronization of a pair of \emph{AirPods} and primary to secondary switching.
The \emph{Marconi} Bluetooth chip firmware itself is also based on the \emph{RTKit} framework, as it is just another peripheral.

Note that it is very complex to debug root causes for crashes on the \emph{AirPods}.
As they are an embedded device, they reboot within approximately \SI{2}{\second}.
Thus, a Bluetooth connection reset is indistinguishable from a device reboot when performing wireless tests.

\subsection{MagicPairing Messages}
\label{ssec:messages}

\begin{figure*}[!t]
%TODO maybe use scalebox with 0.8 to make font as small as in the tikz figures?
%even adjustbox doesn't work, it removes a table...

    \begin{subfigure}[b]{\textwidth}

	\center
	\includegraphics[scale=0.8]{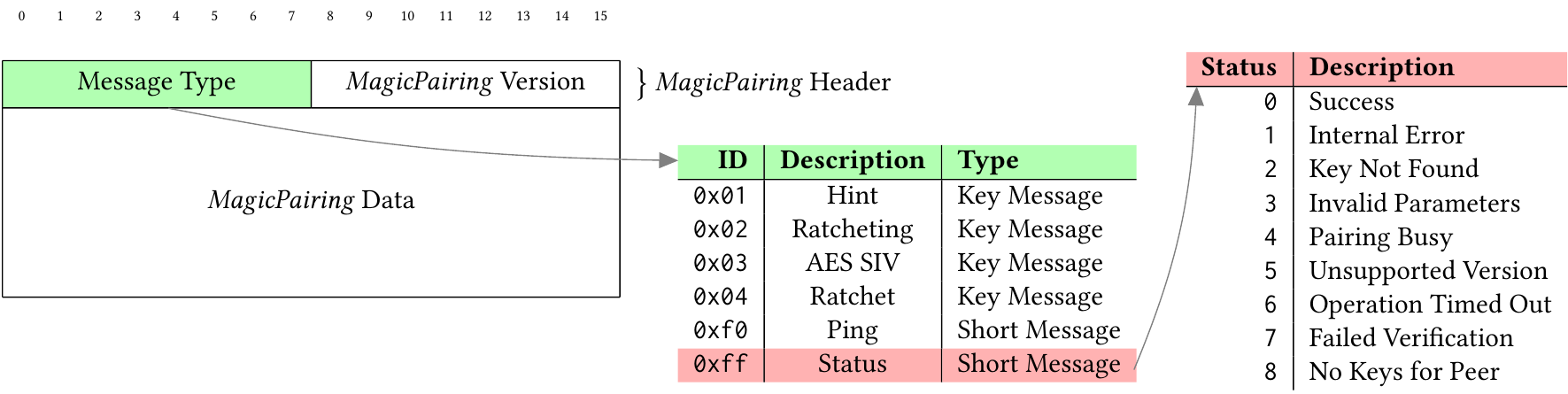}

    \caption{\emph{MagicPairing} message format.} 
    \label{fig:mpFormat}
    \end{subfigure}    
    \begin{subfigure}[b]{\textwidth}
%    \vspace{1em}
%    \begin{bytefield}[bitwidth=1.4em]{16}
%        \bitheader{0-23} \\ \\
%            \bitbox{8}{Num Entries} & \colorbox{blue!30}{\bitbox{16}{Key Type}} \\ 
%            \bitbox{16}{Key Length} & \bitbox[tlr]{8}{} \\
%            \bitbox[lr]{24}{} \\
%            \bitbox[lr]{24}{Key (variable length)} \\
%            \bitbox[blr]{24}{}
%    \end{bytefield}\hspace{4em}\begin{tabular}{ r|l }
%    \rowcolor{blue!30}
%        \textbf{Key Type} & \textbf{Description} \\
%        \hline
%        \texttt{0x0010} & Hint \\
%        \texttt{0x0020} & Nonce \\
%        \texttt{0x0040} & \textit{Unknown}\\
%        \texttt{0x0080} & AES SIV \\
%        \texttt{0x0100} & Ratchet \vspace{8.7em}
%    \end{tabular}
%    
%    \vspace{-15.5em}\hspace{33.5em}\begin{tikzpicture}
%    \path[->,gray] (0,0) edge [bend right=5] (1.3,0);
%	\end{tikzpicture}
%    \vspace{7.5em}
    
    \center
	\includegraphics[scale=0.8]{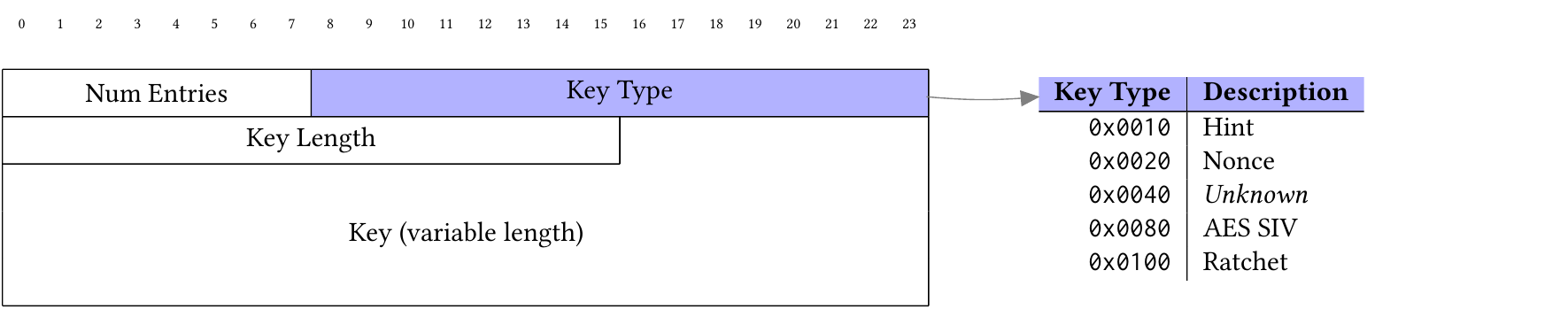}
    \caption{\emph{MagicPairing} key message data.} 
    \label{fig:mpKeyMessageData}
    \end{subfigure}    
    \caption{\emph{MagicPairing} packet formats.}
\end{figure*}

The general layout of a \emph{MagicPairing} message is shown in
\autoref{fig:mpFormat}.  It starts with a \SI{2}{\byte} header, which is followed by data,
depending on the type of the message. In general there are two different types of messages
with a slightly different structure. The first type, a \emph{Key Message}, contains key
material (such as the \emph{AES SIV}, \emph{Ratchet}, or \emph{Hint} data).
The second type, a \emph{Short Message}, contains just
one byte of data after the header. The data in the \emph{Key Message} is in a \gls{TLV}
structure, as shown in \autoref{fig:mpKeyMessageData}. The number of keys is encoded after the header. The \emph{Short Message} contains a fixed amount
of data after the header.

The \emph{MagicPairing} \emph{Ping} message can initiate the protocol. When a device
receives a \emph{Ping} message, it replies with a \emph{Hint} message. While the \emph{Ping}
message does not necessarily need any additional data, it is still \SI{3}{\byte} with
the data set to \texttt{0x00}. The \emph{Status} message indicates success or, in case
of an error, the reason for failing. The \emph{Ratchet} type message seems to be
currently unused, as its reception handler implementation is empty on \emph{iOS} and \emph{macOS}
\texttt{bluetoothd}.

\subsection{MagicPairing AirPods Advertisements}
\label{ssec:advertisement}

In addition to the pairing mechanism provided by \emph{MagicPairing}, it also offers the
capability to decrypt \gls{BLE} advertisements sent by the \emph{AirPods}.
These advertisements have been shown to be linkable to \emph{AirPods} in general~\cite{martin2019handoff}.
Advertisements notify other \emph{Apple} devices of the presence of the \emph{AirPods} and encode battery state information.
When an \emph{iOS} device
receives advertisements for a pair of \emph{AirPods} that belong to the same \emph{Apple} ID as
the \emph{iOS} device, a pop-up shows an \emph{AirPod} image, the name of the \emph{AirPods}, and the current
battery state. 
%The advertisement data is divided into a plaintext part and a ciphertext part. There are
%two types of \emph{MagicPairing} advertisement data, that only differ in the length of the
%plaintext component in the advertisement. In both types, the latter
The encrypted part constitutes the
 \emph{MagicPairing} data. A new key is introduced, which is called
\emph{MagicPairing EncryptionKey}.

%\begin{figure}[!b]
%%    \begin{subfigure}[b]{\columnwidth}
%%	\includegraphics[width=\columnwidth]{../pics/macos_15_3_uplaod.png}
%%	\caption{\emph{macOS 15.3} \texttt{bluetoothd} upload spelling.}
%%	\label{fig:macosUploadSpelling}
%%	\end{subfigure}
%	
%    \begin{subfigure}[b]{\columnwidth}
%	\includegraphics[width=\columnwidth]{../thesis/pics/macos_15_3_rachete.png}
%	\caption{\emph{macOS 15.3} \texttt{bluetoothd} ratchet spelling.}
%	\label{fig:macosRatchetSpelling}
%	\end{subfigure}
%	
%    \begin{subfigure}[b]{\columnwidth}
%	\includegraphics[width=\columnwidth]{../thesis/pics/ios_13_3_rachet.png}
%	\caption{\emph{iOS 13.3} \texttt{bluetoothd} ratchet spelling.}
%	\label{fig:iosRatchetSpelling}
%	\end{subfigure}
%	\caption{Spelling mistakes in \emph{iOS} and \emph{macOS} \texttt{bluetoothd}.}
%	\label{fig:spelling}
%\end{figure}

\subsection{Code Quality}
\label{ssec:spelling}
The \emph{MagicPairing} implementations on
\emph{iOS} and \emph{macOS} contain various spelling mistakes in logging messages, and in
case of the \emph{macOS} \path{bluetoothd} also in function names.
%Our assumption
%likely proved right. The flaws we found regarding the NULL pointer dereferences
%are small coding mistakes that should have been caught during code review, similar to the
%spelling mistakes.
%Given the attack surface of the protocol a thorough review should have
%certainly be done before releasing the code in production.
 For example, the words \emph{Ratchet} and \emph{Upload} were spelled differently various times.
%An excerpt of these spelling
%mistakes is shown in
%\autoref{fig:spelling}.
As these mistakes vary with the stack, each stack was probably implemented by a different developer. While
spelling mistakes are not directly related to flaws in an implementation, they
 leave the impression the code was not extensively reviewed, and development probably outsourced.

%!TEX root = ../magic.tex

\newpage
%TODO fixes layout

\section{Fuzzing with ToothPicker}
\label{sec:fuzzing}
The wireless attack surface of \emph{MagicPairing} is rather large.
First of all, it is available prior pairing---it provides a connection via the \ac{L2CAP}, which is used for all kinds of data transfer within Bluetooth~\cite[p. 252]{bt52}.
Second, the \emph{MagicPairing} attack surface is further enlarged by the different implementations for \emph{iOS}, \emph{macOS}, \emph{RTKit}.
Instead of using a common library, the \emph{macOS} implementation is written in \emph{Objective C}, the \emph{iOS} implementation is based on \emph{C/C++}, and the \emph{RTKit} firmware on the \emph{AirPods} is a slightly feature-restricted variant written in \emph{C}~\cite{ttdennis}.
Last, \emph{MagicPairing} is always available on all \emph{Apple} devices with Bluetooth enabled, no matter if the user owns \emph{AirPods}.

Based on our knowledge about \emph{MagicPairing} and its implementations, we perform further tests.
We implement both, a generic over-the-air fuzzer (\autoref{ssec:fuzzota}) and an \emph{iOS} in-process fuzzer (\autoref{ssec:fuzzprocess}).
While the over-the-air fuzzer is platform-independent and required to confirm vulnerabilities, it is limited in speed and does not provide coverage. 
In contrast, the \emph{iOS} in-process fuzzer is faster and not limited by connection resets, but needs a lot of platform-specific tuning.
Our overall setup is explained in \autoref{ssec:setup}.
As we apply a rather specific tooling to enable \emph{iOS} in-process fuzzing with \frida, we further describe it in \autoref{ssec:frida}

\subsection{Over-the-Air Fuzzing}
\label{ssec:fuzzota}

An over-the-air fuzzer runs independently of the target system. Still, the protocol needs to be re-implemented to fuzz inputs.
Our fuzzer extends \emph{InternalBlue}, which already provides a generic interface to add custom protocols on top of existing Bluetooth stacks, including \emph{iOS} and \emph{macOS}~\cite{mantz2019internalblue}.
This approach has two main advantages that cannot be reached with in-process fuzzing.

\begin{description}
    \item[(+) Platform Independence] The fuzzer is independent of the target
        device's operating system or Bluetooth stack. 
    \item[(+) Few False Positives] The fuzzer behaves just as any
        other Bluetooth peripheral. Anything found can be used comparably easy for a \ac{PoC}.
\end{description}

However, wireless Bluetooth fuzzing has various limitations that motivate us to also perform in-process fuzzing.

\begin{description}
    \item[(-) Connection Termination] The connection is terminated once a few invalid packets are received. Thus, a lot of time is spent on reconnecting to the target. Moreover, it is difficult to nearly impossible to distinguish between a terminated connection and a crashed Bluetooth daemon. % without further debugging.
    \item[(-) Speed] The fuzzer's speed is limited by the physical connection to the target.
    \item[(-) Coverage] Without collecting information from the target, the input cannot be adapted to trigger missing code paths.
\end{description}

As the \emph{MagicPairing} protocol has a low complexity, we implement a generation-based fuzzer, which generates all possible message types.
It randomly generates valid and invalid messages based on the reverse-engineered protocol definition.
Apart from connecting to the target device, no further setup is required for fixed \ac{L2CAP} channels.
%---enabling a rather simple \emph{InternalBlue} fuzzing extension.
%Fixed \ac{L2CAP} channels do not require an authenticated connection on
%\emph{iOS}.
The fuzzer keeps sending the generated \gls{L2CAP} payloads until it receives an
\path{HCI_Disconnection_Complete} event (see \cite[p. 2296]{bt52}). This indicates that the
target device either disconnected due to multiple invalid received messages or due to a
crash. The fuzzer then tries to reconnect to the device.

The target device is
additionally monitored using the \emph{PacketLogger}, which is available for \emph{macOS} and also on mobile devices since \emph{iOS 13} with a \emph{Bluetooth Profile}~\cite{profileslogs}. This enables us to determine if the device crashed and when the connection was terminated. A crash can be
detected by searching for the message \textit{``Connection to the iOS device has been
lost.''}
%The implementation of the \emph{MagicPairing} \gls{L2CAP} fuzzer can be found in
%\cref{app:mp-fuzzer}.

In practice, the \path{HCI_Disconnection_Complete} event is quite unreliable.
In multiple occasions the connection was terminated, but the fuzzer did not receive the
event. This lowers the efficiency of the fuzzer as it needs to estimate
when a connection is terminated in case it did not receive the disconnection event.
Moreover, to reliably send packets and confirm events, we restricted the fuzzer speed to \SIrange{1}{2}{packets\per\second}.
Despite the mentioned issues, the fuzzer ended up finding multiple bugs in the protocol
implementations.

\begin{figure*}[!b]
\center
	\begin{tikzpicture}[minimum height=0.55cm, scale=0.8, every node/.style={scale=0.8}, node distance=0.7cm]
	
	% Devices 
    \node[inner sep=0pt] (iphone) at (-5.5,0)
    {\includegraphics[height=1.5cm]{pics/iphone.pdf}};  
    \node[inner sep=0pt] (iphonex) at (-5.5,0)
    {\includegraphics[height=1.2cm]{pics/iosbt.png}};  
    \node[inner sep=0pt] (laptop) at (5.5,0)
    {\includegraphics[height=1.5cm]{pics/laptop.pdf}}; 
    \node[inner sep=0pt] (tp) at (0,-2)
    {\includegraphics[height=2.1cm]{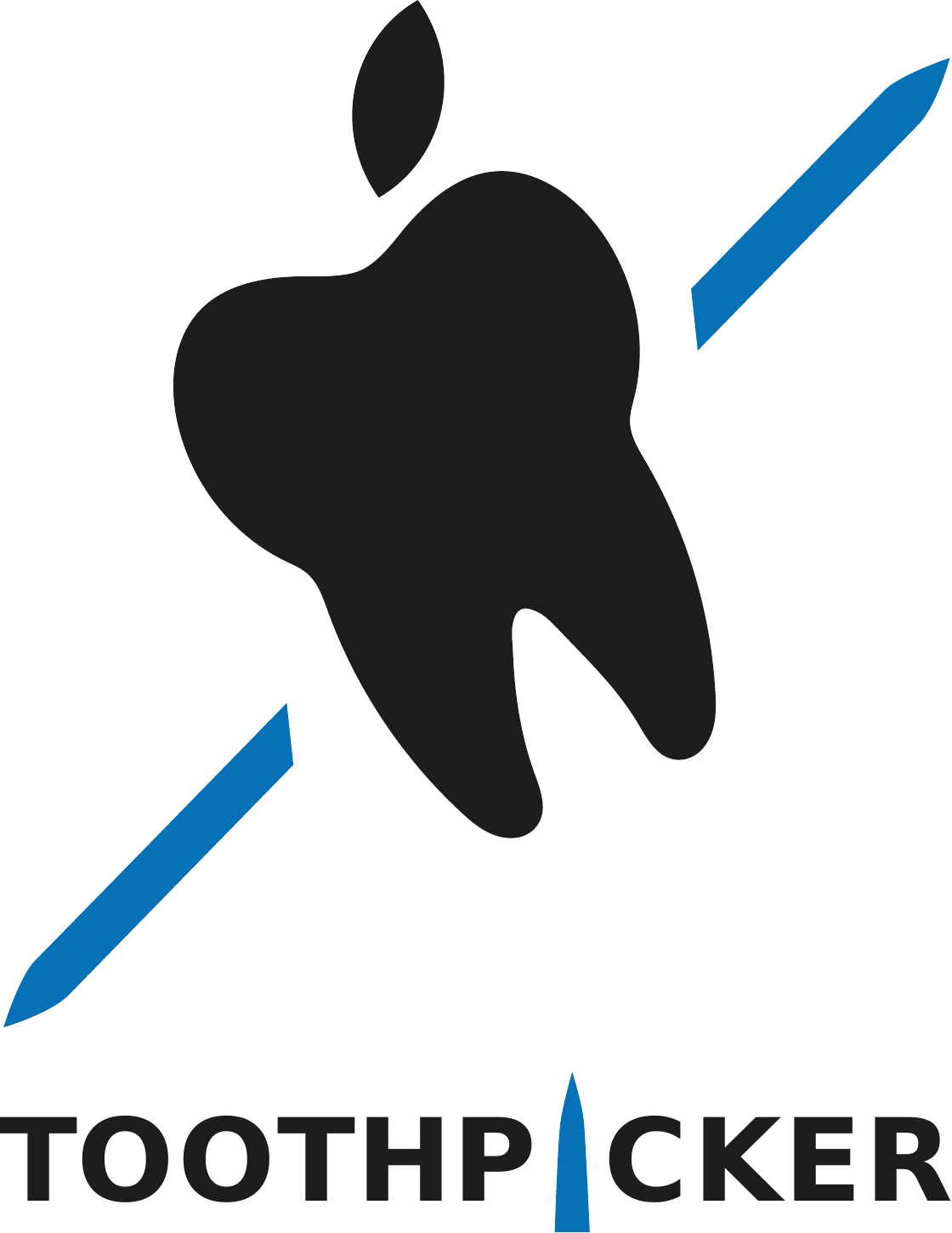}}; 

    % Device Labels
    \node[below=of iphone.south, anchor=south, yshift=0.3cm] (iphonetxt) {iPhone};
    \node[below=of laptop.south, anchor=south, yshift=0.3cm] (laptoptxt) {Laptop};
    
    %\filldraw[align=left, fill=blue!30](-7,0) rectangle node (host) {Broadcom Bluetooth Chip} ++(5,0.5);
    
   \path[->] (-4.4,0.3) edge node[sloped, anchor=center, thick, above] {Coverage, Exceptions} (4.2,0.3);
   \node[] (tcp) at (-0.1, 0) {\textcolor{gray}{TCP/usbmux}};
   \path[<-] (-4.4,-0.3) edge node[sloped, anchor=center, thick, below] {Input} (4.2,-0.3);

    \filldraw[align=center, fill=white](-7.5,-2) rectangle node (fuzz1) {General Fuzzing Harness} ++(4,0.5);
    \filldraw[align=center, fill=white](-7.5,-2.5) rectangle node (fuzz2) {Specialized Fuzzing Harness} ++(4,0.5);

    \filldraw[align=center, fill=white](4,-2) rectangle node (mgr) {Manager} ++(3,0.5);
    \filldraw[align=center, fill=white](4,-3) rectangle node (im) {Input Mutation} ++(3,0.5);

    \filldraw[align=center, fill=white, draw=gray](8,-1) rectangle node (co) {Coverage} ++(2,0.5);
    \filldraw[align=center, fill=white, draw=gray](8,-2) rectangle node (cp) {Corpus} ++(2,0.5);
    \filldraw[align=center, fill=white, draw=gray](8,-3) rectangle node (cr) {Crashes} ++(2,0.5);

    \filldraw[align=center, fill=white](11,-3) rectangle node (ota) {OTA Fuzzer} ++(3,0.5);
    
    %\path[<->] (mgr) edge node[thick] {} (im);
    %\path[<->] (3,-1.75) edge node[thick] {} (4,-1.75);
	\path[<->] (5.5,-1.95) edge node[thick] {} (5.5,-2.55);     % mgr input mut
    \path[<->] (7,-1.6) edge node[thick] {} (8,-0.75);
    \path[<->] (7,-1.75) edge node[thick] {} (8,-1.75);
    \path[<->] (7,-1.9) edge node[thick] {} (8,-2.75);
    \path[->] (10,-2.75) edge node[thick] {} (11,-2.75);

    \end{tikzpicture}
    %\vspace{-1.5em} %TODO looks okay here
\caption{\emph{ToothPicker} fuzzing setup.}
\label{fig:fuzzsetup}
\end{figure*}

\subsection{In-Process Fuzzing}
\label{ssec:fuzzprocess}

%The over-the-air fuzzing approach shows two major drawbacks. First, it depends on a
%physical connection that is not reliable under the load of the fuzzing input and requires
%frequent reconnects. Second, there is no feedback about coverage that is achieved with
%the generated inputs.

An in-process,
coverage-guided fuzzer improves the efficiency when fuzzing the reception handlers of interesting
\gls{L2CAP}-based protocols. While the throughput of messages and the stability of
the payload delivery is much higher than in the over-the-air implementation, the
in-process fuzzer comes with a different set of drawbacks. 

\begin{description}
    \item[(-) Many False Positives] The usual operation of the Bluetooth daemon is
        altered, which can lead to unexpected behavior or crashes that are related to the fuzzing operation itself.
    \item[(-) Platform Dependence] Injecting and preparing the fuzzer
        inside the target process differs significantly for different operating systems.
        Even within the same Bluetooth stack, function addresses and implementation details change with updated versions and need to be adapted.
\end{description}
 
We reduce the false positives by minimizing the amount of crashes related to the
injected fuzzing code. This is done by observing any side-effects during
fuzzing and patching the affected functions.
%Despite all the advantages of the in-process fuzzing method, the over-the-air implementation
%is still required to verify the identified crashes.

\emph{RTKit} currently cannot be altered, thus, only the \emph{macOS} and jailbroken \emph{iOS} Bluetooth stack remain for in-process fuzzing. As \emph{iOS} jailbreaks were comparably rare in the past but became available with \emph{checkm8} and \emph{checkra1n} recently~\cite{checkra1n}, we implement an in-process fuzzer for \emph{iOS}.

In practice, the \emph{iOS} in-process fuzzer's speed varies between \SIrange{5}{30}{packets\per\second}. Though, as connections are not dropped with in-process fuzzing, the overall speedup is much higher.

\subsection{Setup Overview}
\label{ssec:setup}

\autoref{fig:fuzzsetup} shows the fuzzing setup, with a main focus on the specialized in-process fuzzer.
The in-process fuzzer is divided into two components: \emph{(1)} The manager running on a computer, and \emph{(2)} the fuzzing harness running on the target device.

The manger starts and maintains the fuzzing process. It injects fuzzing harness into the target process and handles the communication with it. Additionally, it maintains a set of crashes occurred during fuzzing, a corpus to derive inputs, and coverage information collected during fuzzing. The manager generates new inputs by sending entries to the corpus of the input mutation component, which randomly mutates the input based on a seed.

The fuzzing harness is divided into two sub-components. The first component is a general fuzzing harness, which is responsible for the overall fuzzing of \texttt{bluetoothd}. It creates virtual connections and applies patches ensuring a stable fuzzing process. Moreover, it collects code coverage and receives fuzzing input from the manager. The second component, the specialized fuzzing harness, is specific for the target function and protocol to be fuzzed, such as \emph{MagicPairing}. It is responsible for preparing the received input and calling the function handler, as well as any other preparation needed to fuzz the protocol-specific reception handler function.

The fuzzer is initialized with an initial corpus of valid protocol messages, i.e., function arguments.
It then collects the initial coverage by sending the initial corpus to the fuzzing harness. The specialized harness
executes the payloads. The collected coverage is returned to the manager.

Once the initial coverage is collected, the actual fuzzing begins. %In each iteration, multiple steps happen.
The manager picks one of the entries in the corpus and a seed value. These are passed to the input mutator, which mutates the input and sends it back to the manager. The manager sends the mutated input to the specialized fuzzing harness. If desired, the specialized fuzzing harness further mutates the input---which is required for fields that require deterministic values or length fields.
In this case, the specialized fuzzing harness first reports the modified input back to the manager before calling the function under test.
This ensures that the additional mutation is saved, even when the injected harness crashes together with the target.
While the function is called, the harness collects basic block coverage. There are three possible results of the function call:

\begin{description}
\item[Ordinary Return] The function was executed successfully and returns. The collected coverage is reported to the manager.
\item[Exception] The function results in an exception, which is returned to the manager. The manager stores the input and the exception as a crash.
\item[Uncontrolled Crash] The target, i.e., \texttt{bluetoothd}, crashes in a thread not controlled by the fuzzing harness. It crashes and generates a crash report. In this case, the exception cannot be sent to the manger. However, the manager detects this crash and stores the generated input as a crash. The corresponding crash report is manually gathered from the operating system.
\end{description}

These results may contain false positives, even in the case of an exception.
Therefore, we verify identified crashes with the over-the-air fuzzer.

\subsection{Attaching the In-Process Fuzzer}
\label{ssec:frida}
Our in-process fuzzer is based on \texttt{frizzer}~\cite{frizzer}, which provides a basic fuzzing architecture including coverage collection, corpus handling, and input mutation. These are already a large part of the manager component. Our fuzzer, like \texttt{frizzer}, is built on \frida, which is a dynamic instrumentation toolkit~\cite{frida}. \frida can inject code into a target process using \emph{JavaScript}.
Thus, our fuzzing harness is implemented in \emph{JavaScript} and injected into \texttt{bluetoothd}.
The manager is implemented in \emph{Python}, as \frida also provides \emph{Python} bindings.
We use the test case generator \texttt{radamsa} as input generator component~\cite{radamsa}.

On \emph{iOS}, \texttt{bluetoothd} is missing symbols.
Nonetheless, we can identify various functions by static reverse engineering. These include creating a \ac{BLE} handle, or creating an \ac{ACL} handle, which is needed to receive \ac{L2CAP} data.
Due to the lack of symbols, we need to resolve function pointers via their static
offsets to make them callable with \frida. In the following example, these offsets are valid for an \emph{iPhone 7} on \emph{iOS 13.3}. In \autoref{lst:forgeACLfrida}, we call the function that creates an \ac{ACL} handle. The input arguments are a Bluetooth address, and another state value set to \texttt{0} as found by dynamic analysis.
Similar to calling the \ac{ACL} connection creation function, we also call the specialized \emph{MagicPairing} handler.

Even fake connections created as in \autoref{lst:forgeACLfrida} can be disconnected. We
keep the connection alive by overwriting the function \path{OI_HCI_ReleaseConnection},
named according to the debug strings.
Hooking and replacing such functions prevents connection structures from being destroyed.

Note that this in-process fuzzing disconnection prevention does not work for over-the-air fuzzing. When a connection is initiated by the Bluetooth chip itself, it holds an \ac{HCI} handle, which the stack uses to reference the connection. Moreover, the chip holds additional state to keep the connection alive. While we can control \texttt{bluetoothd} with \frida hooks, we cannot overwrite chip-internal behavior.

\begin{figure}[!b]
\begin{lstlisting}[caption={Creating a forged ACL handle using \protect\frida.}, label=lst:forgeACLfrida]
// Create a buffer for the Bluetooth address
var bd_addr = Memory.alloc(6);
// Resolve function address
var base = Module.getBaseAddress("bluetoothd");
var fn_addr = base.add(0xc81a0);  // iOS 13.3, iPhone 7
// Create JavaScript-callable function reference
var allocateACLConnection = new NativeFunction(fn_addr,  "pointer", ["pointer","char"]);
// Write a (random) Bluetooth address to memory
bd_addr.writeByteArray([0xca,0xfe,0xba,0xbe,0x13,0x37]);
// Call the function and create a forged ACL connection
var handle = allocateACLConnection(bd_addr, 0);
\end{lstlisting}
\end{figure}

\begin{figure}[!b]
\vspace{-1em}
\begin{lstlisting}[caption=Excerpt of an \emph{MP1}-related crash log.,label=lst:crashlog-mp]
Exception Type:  EXC_BAD_ACCESS (SIGSEGV)
Exception Subtype: KERN_INVALID_ADDRESS at @\textcolor{darkred}{0x00000000000000a8}@
VM Region Info: @\mbox{\textcolor{darkred}{0xa8 is not in any region.}}@ Bytes before following region: 4298293080
[...]
Termination Signal: Segmentation fault: 11
Termination Reason: Namespace SIGNAL, Code 0xb
Terminating Process: exc handler [958]
\end{lstlisting}
\end{figure}

%!TEX root = ../magic.tex

\newpage
%TODO fixes layout

\section{Vulnerabilities in the MagicPairing Implementations}
\label{sec:vulns}

%\todo[color=blue!30]{Reviewer A: Did you also fuzz the TLV fields? Those are typically prone to errors.}

In the following, the identified vulnerabilities in the \emph{MagicPairing} protocol are
described. All vulnerabilities are summarized in \autoref{tbl:mp-vulns}.

\begin{table*}
    \caption{List of identified \emph{MagicPairing} and L2CAP vulnerabilities, status April 28 2020.}
    \begin{tabular}{ l|l|l|l|l|r|l }
        \textbf{ID} & \textbf{Attack} & \textbf{Effect} & \textbf{Detection Method} & \textbf{OS} & \textbf{Disclosure} & \textbf{Status} \\
        \hline
        {MP1} & {Ratcheting} & Crash & Over-the-Air, In-Process & iOS & Oct 30 2019 & Not fixed \\
        {MP2} & {Hint} & Crash & Over-the-Air, In-Process & iOS & Dec 4 2019 & Not fixed \\
        {MP3} & {Ratcheting} & Crash & Over-the-Air & macOS & Oct 30 2019 & Not fixed \\
        {MP4} & {Hint} & Crash & Over-the-Air & macOS & Oct 30 2019 & Not fixed \\
        {MP5} & {Ratcheting} & Crash & In-Process & iOS & Mar 13 2020 & Not fixed \\
        {MP6} & {Ratcheting Abort} & Crash & In-Process & iOS & Mar 13 2020 & Not fixed \\
        {MP7} & {Ratcheting Loop} & \SI{100}{\percent} CPU Load & Over-the-Air & macOS & Oct 30 2019 & Not fixed \\
        {MP8} & {Pairing Lockout} & Disassociation & Manual & iOS \& macOS & Feb 16 2020 & Not fixed \\
        {L2CAP1} & {L2CAP Zero-Length} & Crash & Over-the-Air & RTKit & Dec 4 2019 & Not fixed \\
        {L2CAP2} & {L2CAP Groups} & Crash & In-Process & iOS 5--13 & Mar 13 2020 & Not fixed \\
    \end{tabular}
    \label{tbl:mp-vulns}
\end{table*}

\subsection{Null Pointer Dereferences}
Testing the \emph{MagicPairing} protocol resulted in multiple \texttt{NULL} pointer dereferences 
or dereferencing addresses in the \texttt{NULL} page. The \texttt{NULL} page is not mapped on \SI{64}{\bit} \emph{iOS} and \emph{macOS}.
This results in a \texttt{bluetoothd} crash.
\texttt{launchd} immediately restarts \texttt{bluetoothd} after crashing. 
Thus, these bugs are merely a \texttt{bluetoothd} \ac{DoS}. An attacker does not have any control over the dereferenced value, and we assume that these dereferences are not exploitable.

\begin{figure}[!b]
\begin{lstlisting}[language=C,caption=Pointer dereference  \emph{MP1}.,label=lst:pseude-deref]
void recv_mp_ratchet_aes_siv(char *bd_addr, char *data) {
    [...]
    // Returns NULL for unknown Bluetooth addresses
    mp_entry = lookup_mp_entry_by_bd_addr(bd_addr);
    [...]
    // The NULL entry is dereferenced with an offset
    memmove(mp_entry->remoteAESSIV, data + aessiv_offset, 0x36);
}
\end{lstlisting}
\end{figure}

\subsection{MP1: iOS Ratcheting}
When sending
a \emph{MagicPairing} \emph{Ping} message to an \emph{iOS} device from a Bluetooth device that is
not a known pair of \emph{AirPods}, it responds that it does not
have a hint for this sending device. If a \emph{Ratcheting} message is then sent to
the device, \texttt{bluetoothd} will crash while trying to dereference a pointer in the \texttt{NULL}
page. \autoref{lst:crashlog-mp} shows an excerpt of the crash log that is generated by the
operating system.

The invalid access to address \texttt{0xa8} is caused by a missing check for the
return value of a lookup function shown in \autoref{lst:pseude-deref}. The function looks up an entry in \texttt{bluetoothd}'s
table of known \emph{MagicPairing} devices by the sender's Bluetooth address and returns \texttt{NULL}. The issue is that
this return value is never checked and assumed to be a pointer to a valid
\emph{MagicPairing}-related structure. Then, to respond to the \emph{Ratcheting}
message, the structure is accessed at offset \texttt{0xa8}, which leads to the crash.

\subsection{MP2--5: macOS/iOS Hint and Ratcheting}
\emph{MP2--5} have a similar cause as the previous dereference in \emph{MP1}. The return value of the
lookup function is not properly verified. On \emph{iOS} and \emph{macOS}, this affects the
\emph{Ratcheting} \emph{(MP1, MP3, MP5)} and
the \emph{Hint} \emph{(MP2, MP4)} messages.
As before, they lead to a dereference of an invalid
address, which is a fixed offset into a \emph{MagicPairing} structure at address
\texttt{0x0}. Thus, all vulnerabilities are equally unlikely exploitable other than
crashing \texttt{bluetoothd}. The reason why the \emph{Ratcheting} messages lead to
different crashes on \emph{iOS} is that the order of keys in the message determines which fields
in the \texttt{mp\_entry} are accessed.

\subsection{MP6: Ratcheting Abort}
This crash is caused by an assertion failure that leads to an \texttt{abort}. The code
that parses the \emph{Ratcheting} message attempts reading from the message buffer. An assertion ensures that it does not read beyond this buffer.
However, if the assertion fails, the parser does not return gracefully and instead calls
\texttt{abort}, which leads to the termination of \texttt{bluetoothd}.

\subsection{MP7: Ratcheting Loop}
The \emph{macOS}
\texttt{bluetoothd} can be forced to enter a ratcheting loop with a very large iteration count.
Unlike the previous vulnerabilities, this issue is not solely caused by implementation
mistakes, but originates from an inherent problem in the protocol's design. The receiver trusts the values sent in the \emph{Hint} message, without
verifying that it was actually sent by a known \emph{MagicPairing} peer. An attacker can
forge the \emph{Ratchet} value in the \emph{Hint} message.
The \emph{Hint} message also includes a nonce, but this is random. The \emph{Hint} value itself, which is encrypted and could be used to verify the sender's Bluetooth address, is ignored. Instead, \emph{macOS} trusts the connection's Bluetooth address.

 Setting the \emph{Ratchet} to a
very high value will cause \texttt{bluetoothd} to enter a long ratcheting loop. The \emph{Ratchet}
field holds a \SI{4}{\byte} value, thus the maximum value of a \emph{Ratchet} can be
\texttt{0xffffffff}. During normal usage however, the \emph{Ratchet} is only incremented
for every pairing process. Therefore, it is rather small in practice. The attack was
tested on a \emph{MacBook Pro Early 2015, 13-inch, 2.9 GHz Dual Core i5} on \emph{macOS Catalina 10.15}
with an initial \emph{Ratchet} value of \texttt{2}. Sending a \emph{Hint} message with a
\emph{Ratchet} value of \texttt{0xffffffff} caused \texttt{bluetoothd} to enter a
ratcheting loop, with the local \emph{Ratchet} value increasing at a rate of approximately \SI{7000}{\per\second}---causing a ratcheting loop running multiple days.

During the ratcheting loop attack, the \texttt{bluetoothd} reception thread is blocked. This disables further Bluetooth-based communication, for example, the device under attack can no longer receive files via \emph{AirDrop}.

%\begin{figure}[!b]
%\begin{lstlisting}[caption=\emph{AirPod} ratcheting loop., label=lst:ratchet-threshold]
%if localRatchet < remoteRatchet - 9:
%    sendMPStatus(1)
%else
%    ratchetingLoop()
%\end{lstlisting}
%\end{figure}

\begin{figure}[!b]

\center

	\begin{center}
	\begin{tikzpicture}[minimum height=0.55cm, scale=0.8, every node/.style={scale=0.8}, node distance=0.7cm]
	
	% Devices 
    \node[inner sep=0pt] (iphonea) at (3,0)
    {\includegraphics[height=1.5cm]{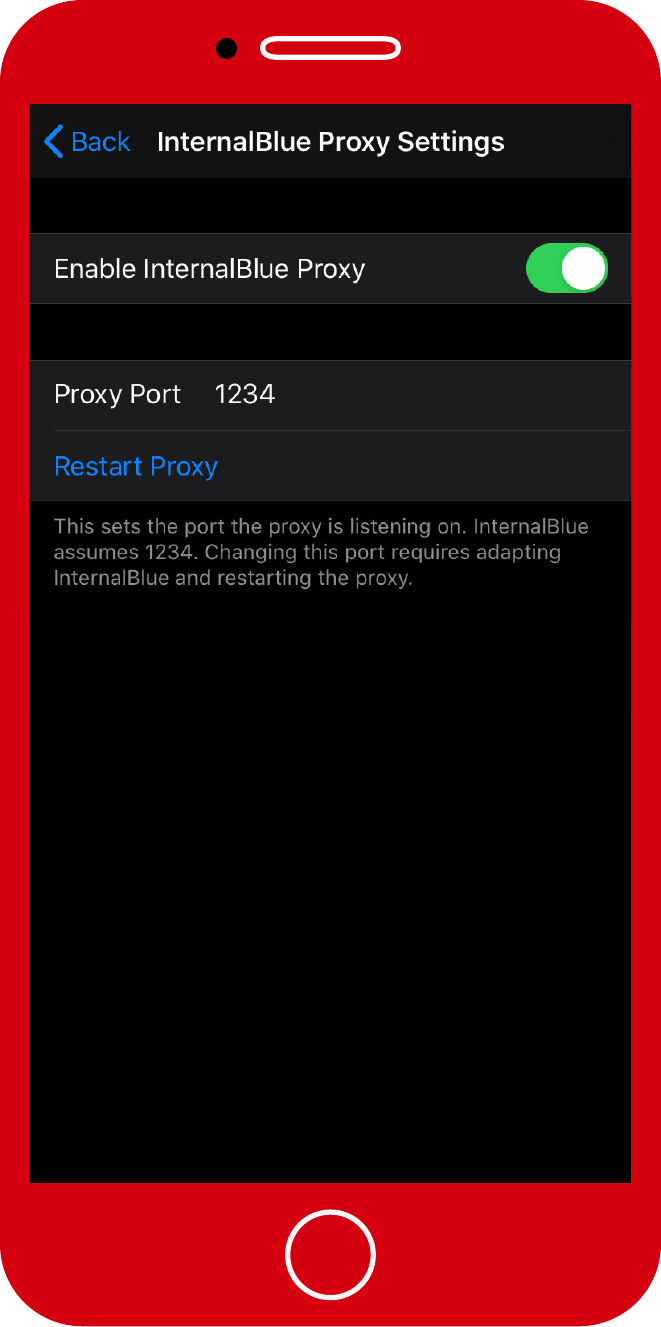}}; 
    \node[inner sep=0pt] (airpoda) at (3,-2.5)
    {\includegraphics[height=1.5cm]{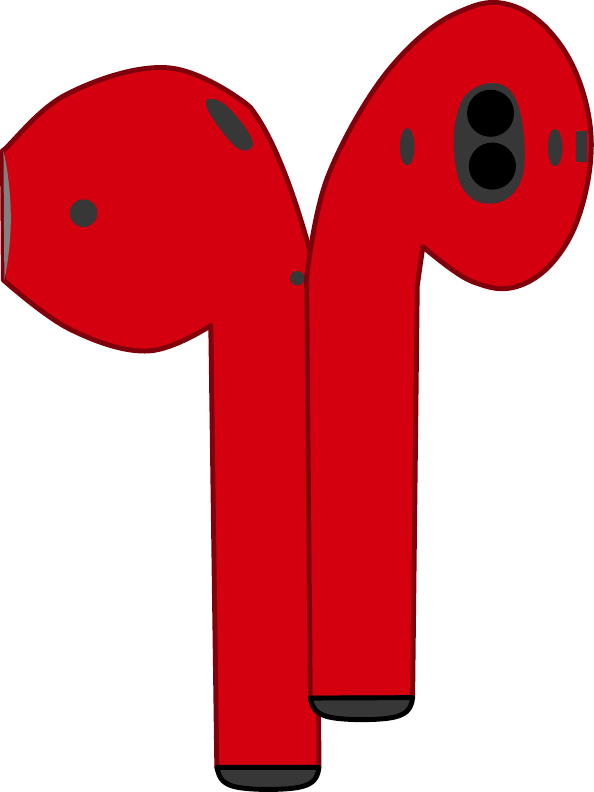}}; 
    \node[inner sep=0pt] (iphone) at (9,0)
    {\includegraphics[height=1.5cm]{pics/iphone.pdf}}; 
    \node[inner sep=0pt] (iphonex) at (9,0)
    {\includegraphics[height=1.2cm]{pics/iosbt.png}};   
    \node[inner sep=0pt] (laptop) at (10,0)
    {\includegraphics[height=1.5cm]{pics/laptop.pdf}};

    % Device Labels
    \node[below=of iphone.south, anchor=south, yshift=0.3cm,xshift=2em] (iphonetxt) {iPhone/MacBook};
    \node[below=of iphonea.south, anchor=south, yshift=0.3cm] (iphonetxt) {\emph{InternalBlue} iPhone};
    \node[below=of airpoda.south, anchor=south, yshift=0.3cm] (airpodtxt) {Fake AirPods};

    % Device Lines
    \draw[-,gray,dashed,thick] (3,-1.3) -- (3,-1.7);
    \draw[-,gray,dashed,thick] (3,-3.8) -- (3,-6.5);
    \draw[-,gray,dashed,thick] (9.5,-1.3) -- (9.5,-6.5);

    \node[align=left,left] at (2.8, -1.5) {\textcircled{1} Set \texttt{BdAddr}};

    \path[->] (3.2,-4.3) edge node[sloped, anchor=center, above, text width=5cm,yshift=-0em] {\textcircled{2} Ping} (9.3,-4.3);
    
    \path[<-] (3.2,-5) edge node[sloped, anchor=center, above, text width=5cm,yshift=-1.25em] {\textcircled{3} Hint \\ \textcolor{gray}{\texttt{ratchet\_host}}} (9.3,-5);

    \path[->] (3.2,-6) edge node[sloped, anchor=center, above, text width=5cm,yshift=-1.25em] {\textcircled{4} Ratcheting \\ \textcolor{gray}{\texttt{ratchet\_airpod=ratchet\_host + 10}}} (9.3,-6);

    \node[align=left,right,darkred] at (9.7, -6.2) {\textcircled{5} Ratchet \\ \hspace{1.5em}Lockout};

	\end{tikzpicture}
	\end{center}
	
\caption{Lockout attack.}
\label{fig:lockout}
\end{figure}

%\todo{alex: Interessant fände ich auch, ob sich der Angriff automatisieren lassen würde, wenn man die BLE advertisements der AirPods beobachtet. Dann würde man ja eine Bluetooth Adresse bekommen, die man dann im Angriff verwenden könnte.
%Die iOS Bluetooth Adresse könnte man genauso bekommen, wie die der AirPods.\\
%PDF annotation:\\
%can this attack be automated? wait for ble advertisement from airpods to obtain ble address, wait for ble advertisements from ios device in range, start attack with airpods ble for every ios device in range.}

\subsection{MP8: Pairing Lockout}
It is possible to corrupt the established pairing between an \emph{iOS} or \emph{macOS} device and a
pair of \emph{AirPods}. For this, an attacker needs to know the victim's Bluetooth address, as
well as the target \emph{AirPods}' Bluetooth address. The attacker can manipulate the local
ratchet value of a host device by sending one or more \emph{Ratcheting} messages with
a ratchet value higher than the device's current one. The current ratchet value can be
obtained by sending a \emph{Ping} message to the host. It responds with a
\emph{Hint} message, which contains its current local ratchet value. This value can then
be incremented and sent in a \emph{Ratcheting} message. The keys for encrypting the
\emph{AES-SIV} value are not required, as the ratchet value is sent in plaintext.
Therefore, an attacker can set a bogus value for the \emph{AES-SIV} part of the message and set
the incremented ratchet value. Then, the receiving host starts a ratcheting loop.
As \texttt{bluetoothd} on \emph{iOS} has a timeout functionality, the forged ratchet value
should not be chosen too high. Once
the ratcheting loop is finished, the host's local ratchet value is successfully increased,
even if the decryption of the \emph{AES-SIV} entry of the message fails.
This corrupts an active pairing because the \emph{AirPods} have a threshold value
for the discrepancy between their local ratchet value and the value received by the paired
host.
%A pseudo code representation of this check is shown in \autoref{lst:ratchet-threshold}.

This causes the \emph{AirPods} to decline the continuation of the \emph{MagicPairing} protocol
and thus the whole pairing process. The user does not have any options to reset the
\emph{MagicPairing} data and does not get any feedback about the error. The only
solution is to reset the \emph{AirPods} and freshly pair them with the user's \emph{iCloud} account. 

As shown in \autoref{fig:lockout}, the attack can be conducted as follows:

\begin{enumerate}
    \item The attacker changes the Bluetooth address to that of the target's \emph{AirPods}.
    
    \item The attacker connects to the victim and sends a \emph{Ping} message\footnote{The attacker could also send regular \emph{MagicPairing AirPod} advertisements (\autoref{ssec:advertisement}), but they are encrypted and regularly change.}  to initiate a \emph{MagicPairing} process.
    
    \item The victim responds with a \emph{Hint} message which contains its current local
        ratchet value.
    
    \item The attacker increases this value by 10 and sends a \emph{Ratcheting}
        message with the incremented ratchet value and a random \emph{AES-SIV} value.
    
    \item The victim will start the ratcheting loop with the received ratchet value and
        derive the \emph{SIV Key} for decrypting the \emph{AES-SIV} value. As the \emph{AES-SIV} value is
        random, the victim will not be able to decrypt it and sends a \emph{Status}
        message indicating an internal error. However, its local ratchet value stays
        incremented and is not reset to its previous value.
\end{enumerate}

The issue originates from using an untrusted ratchet to increment an
internal value and execute a key rotation. As the ratchet value is neither encrypted nor
authenticated, an attacker can easily forge the ratchet. 

A solution to this problem is to only store the incremented ratchet value and the
rotated key when the \emph{AES-SIV} part of the message was successfully decrypted. Otherwise,
the whole \emph{MagicPairing} message should be considered untrusted and the ratchet value
should stay as it was before.
%\todo{alex: does this mitigation work? how does the siv implementation validate the decryption? how long is the authentication tag? if it short (like for handoff) it can be easily brute forced.}

\subsection{L2CAP1: L2CAP Zero-Length}
While fuzzing \emph{MagicPairing} over-the-air, we identified a crash in the \emph{RTKit}
Bluetooth stack, more specifically, the \emph{AirPods 1} and \emph{2}. When
sending an L2CAP message with the length field set to zero and no payload, the
\emph{AirPods} crash. As there are no publicly documented debugging capabilities for the
\emph{AirPods}, it is not possible to tell whether the Bluetooth thread or the whole operating
system crashes. We observe that the music stops playing, the
connected \emph{iPhone} reports the \emph{AirPods} as disconnected, and after a few
seconds, the \emph{AirPods} play a sound indicating a successful connection.

\subsection{L2CAP2: L2CAP Groups}
This crash is another \texttt{NULL} pointer dereference, albeit more severe than the
previous ones.  It is
accessible via both \ac{BLE} and Classic Bluetooth and is part of \emph{L2CAP Group} feature. This is indicated by logging messages in the crashing function that mention
the file \path{corestack/l2cap/group.c}. However, the \emph{L2CAP Group} feature is no longer
supported since Bluetooth 1.1. We assume the group reception
function has been accidentally left in the code.
In the newest Bluetooth specification, the channel ID \texttt{0x0002} is reserved for connectionless
traffic instead of group traffic~\cite[p. 1035]{bt52}.

Depending on the data that is received, the \emph{L2CAP Group} handler
tries to find a matching entry in a function table allocated on the heap. However, this
table has only been allocated, not initialized. Thus, all its entries are zero. When the
payload starts with a \texttt{NULL} byte, the first entry is identified as matching entry.
The code then tries to jump to the function pointer stored in that table entry, which also is a \texttt{NULL} pointer. However, any control over this table would
immediately result in control over the instruction pointer. In addition to an \emph{iPhone
7} on \emph{iOS 13.3}, we were able to reproduce the crash on an \emph{iPad 2} with
\emph{iOS 9.3.5} (released on August 25 2016), and an \emph{iPhone 4} with \emph{iOS 5.0.1} (released on November 10 2011).
While the crash is not critical per se, it shows how long the \emph{iOS}
Bluetooth stack has not been tested. As \emph{iOS 5} and \emph{9} still had another Bluetooth stack
architecture, the crash is within \texttt{BTServer} instead of \texttt{bluetoothd}.

%!TEX root = ../magic.tex

\section{Conclusion}
\label{sec:conclusion}
In this paper, we showed how \emph{Apple} deals with seamless pairing of Bluetooth
peripherals in their large connected ecosystem. While \emph{MagicPairing} is proprietary,
its general ideas and techniques can be integrated into other \ac{IoT} ecosystems.
Furthermore, other Bluetooth peripheral vendors could benefit from the \emph{MagicPairing}
protocol and infrastructure. All \emph{Apple} needs to do is to provide an \ac{API} that
lets developers generate and receive an \emph{Accessory Key} that is stored in the user's
\emph{iCloud} account. Vendors could then implement \emph{MagicPairing} in their products
and benefit from the same security properties and seamless pairing experience as the
\emph{AirPods}.

\emph{Apple's} three different Bluetooth stacks for \emph{iOS}, \emph{macOS}, and \emph{RTKit} also reflect the variety of Bluetooth implementations outside of their ecosystem. Many vendors choose to implement their own stacks and protocols. This makes efficient testing of Bluetooth devices challenging, but our over-the-air fuzzing setup based on \emph{InternalBlue} can also be useful to test further Bluetooth stacks. As \emph{MagicPairing} is a rather simple protocol, over-the-air fuzzing was sufficient to identify multiple vulnerabilities, despite the lack of speed and coverage information. However, our \emph{iOS}-based in-process fuzzer had a better performance in practice.

Overall, \emph{Apple} keeps their Bluetooth ecosystem rather closed to third-party vendors. Already using Classic Bluetooth requires them to apply for \ac{MFi}. However, this enables an overall smooth user experience. Bluetooth runs silently in the background most of the time and manages tasks like \emph{AirDrop} and \emph{Handoff}~\cite{milan,alexander}. Since \emph{iOS 13}, the Bluetooth icon has been removed from the status bar, even during audio streaming. Any incentive for disabling Bluetooth in the \emph{Apple} ecosystem is missing.

While all of this is great for user experience, we were surprised by the vulnerabilities uncovered within \emph{MagicPairing}. We assume that this protocol never had an extensive code review and was never fuzzed before integrating it as always-active Bluetooth background service. We are looking forward to \emph{Apple} integrating patches for the vulnerabilities we identified, but also hope that they will elaborate their other wireless protocols better in the future.

\begin{acks}
We thank Bianca Mix, Oliver P\"ollny, and Alexander Heinrich for proofreading this paper.
Moreover, we thank Matthias Hollick for his feedback and Anna Stichling for the \emph{ToothPicker} logo.

This work has been funded by the German Federal
Ministry of Education and Research and the Hessen State Ministry for
Higher Education, Research and the Arts within their joint support of
the National Research Center for Applied Cybersecurity ATHENE, as well as by the Deutsche Forschungsgemeinschaft (DFG) -- SFB 1119 -- 236615297.
\end{acks}

%%
%% The next two lines define the bibliography style to be used, and
%% the bibliography file.
\balance
\bibliographystyle{ACM-Reference-Format}
\bibliography{bibfile}
%%
%% If your work has an appendix, this is the place to put it.

\end{document}